\documentclass{iopart}

\usepackage{aas_macros}
\usepackage{xspace}
\usepackage{graphicx}
\usepackage{iopams}

\newcommand{\modl}[1]{model \texttt{#1}}
\newcommand{\modls}[1]{models \texttt{#1}}
\newcommand{\Modl}[1]{Model \texttt{#1}}

\newcommand{\modelname}[1]{\texttt{#1}}

\newcommand{\nusp}{neutrinosphere\xspace}
\newcommand{\nusps}{neutrinospheres\xspace}
\newcommand{\panel}[1]{panel (#1)}
\newcommand{\Panel}[1]{Panel (#1)}
\newcommand{\banel}[1]{{#1}}

\newcommand{\pns}{proto neutron star\xspace}

\newcommand{\ie}{i.e.\xspace}
\newcommand{\cf}{cf.\xspace}
\newcommand{\eg}{e.g.\xspace}
\newcommand{\viz}{viz.\xspace}
\newcommand{\wrt}{w.r.t.\xspace}

\newcommand{\msec}{\textrm{ms}}

\newcommand{\km}{\textrm{km}}
\newcommand{\cm}{\textrm{cm}}

\newcommand{\erg}{\textrm{erg}}

\newcommand{\sek}{\textrm{s}}

\newcommand{\gccm}{\textrm{g\,cm}^{-3}}

\newcommand{\Gauss}{\textrm{G}}

\newcommand{\msol}{M_{\odot}}

\newcommand{\zehn}[1]{10^{#1}}
\newcommand{\zehnh}[2]{{#1} \times 10^{#2}}

\newcommand{\figref}[1]{Fig.\,\ref{#1}}
\newcommand{\tabref}[1]{Tab.\,\ref{#1}}
\newcommand{\secref}[1]{Sect.\,\ref{#1}}

\begin{document}

\title[Magnetic fields and rotation]{Core collapse with magnetic fields and rotation}

\author{M.~Obergaulinger${}^1$, O.~Just${}^{2,3}$, M.A.~Aloy${}^1$}

\address{%%
  ${}^1$ Departamento de Astronom\'{\i}a y Astrof\'{\i}sica,
  Universidad de Valencia,  C/ Dr.~Moliner 50, E-46100 Burjassot,
  Spain; 
  \\ ${}^2$ Astrophysical Big Bang Laboratory, RIKEN Cluster for
  Pioneering Research 2-1 Hirosawa, Wako, Saitama 351-0198, Japan;
  \\ ${}^3$ Max-Planck-Institut f\"ur Astrophysik, Postfach 1317, 85741 Garching, Germany
}

\begin{abstract}
  We study the effects of magnetic fields and rotation on the core
  collapse of a star of an initial mass of $M = 20 \, \msol$ using
  axisymmetric simulations coupling special relativistic
  magnetohydrodynamics, an approximately relativistic gravitational
  potential, and spectral neutrino transport.  We compare models of
  the same core with different, artificially added profiles of
  rotation and magnetic field.  A model with weak field and slow
  rotation does not produce an explosion, while stronger fields and
  fast rotation open the possibility of explosions.  Whereas the
  neutrino luminosities of the exploding models are the same as or
  even less than those of the non-exploding model, magnetic fields
  locally in equipartition with the gas pressure provide a strong
  contribution to the shock revival and the acceleration of bipolar
  outflows.  Among the amplification processes generating such strong
  fields, we find the magneto-rotational instability. However, our
  limited grid resolution allows us to find it only in limited regions
  of the model with the strongest pre-collapse field ($10^{11} \,
  \Gauss$) and fastest rotation.
\end{abstract}

\maketitle
\ioptwocol

\section{Introduction}
\label{Sek:Intro}

The evolution of stars with masses above $ \gtrsim 8 \msol$,
depends on parameters such as the metallicity, the
initial mass, the rotational profile, and the magnetic field of the
star, for which nature offers a wide range of possibilities. Moreover,
it also depends on processes such as convection and other
(hydrodynamic) instabilities, mass loss, and mass transfer within
binary systems, whose complex dynamics increase the spectrum of
evolutionary paths
\cite{Langer__2012__ARAA__Presupernova_Evolution_of_Massive_Single_and_Binary_Stars}.
The resulting diversity of conditions of stellar cores after the end
of their hydrostatic evolution translates into very different
scenarios for the evolution after core bounce.  Besides the action of
neutrino heating in combination with hydrodynamic instabilities,
magnetic fields and rotation may contribute to revive the stalled
shock wave and launch a core-collapse supernova (CCSN) explosion.
Furthermore, shock revival may completely fail in a substantial
fraction of all cores, leading finally to the formation of a black
hole (BH) rather than a neutron star as in most other cases
\cite{Adams_et_al__2017__mnras__ThesearchforfailedsupernovaewiththeLargeBinocularTelescope:constraintsfrom7yrofdata}.

Supernova theory has accounted for this large diversity by, \eg,
studying the standard neutrino-driven mechanism across a wide range of
progenitor masses (for recent work, see \eg,
\cite{OConnor_Ott__2011__apj__BlackHoleFormationinFailingCore-CollapseSupernovae,Ugliano_et_al__2012__apj__Progenitor-explosionConnectionandRemnantBirthMassesforNeutrino-drivenSupernovaeofIron-coreProgenitors,Burrows__2013__ReviewsofModernPhysics__ColloquiumPerspectivesoncore-collapsesupernovatheory,Nakamura_et_al__2015__pasj__Systematicfeaturesofaxisymmetricneutrino-drivencore-collapsesupernovamodelsinmultipleprogenitors,Lentz_et_al__2015__apjl__Three-dimensionalCore-collapseSupernovaSimulatedUsinga15MsunProgenitor,Sukhbold_et_al__2016__apj__Core-collapseSupernovaefrom9to120SolarMassesBasedonNeutrino-poweredExplosions,Mueller_et_al__2017__mnras__Supernovasimulationsfroma3Dprogenitormodel-Impactofperturbationsandevolutionofexplosionproperties})
as well as by considering alternative processes contributing to shock
revival and explosion such as modifications of nuclear physics
\cite{Fischer_et_al__2011__apjs__Core-collapseSupernovaExplosionsTriggeredbyaQuark-HadronPhaseTransitionDuringtheEarlyPost-bouncePhase}.
Rotation and magnetic fields are among the most important such
alternatives.  Similar to the simulations set within the standard
scenario, magneto-rotational models have increased in complexity since
the pioneering works of
\cite{LeBlanc_Wilson__1970__ApJ__MHD_SN,Meier_etal__1976__ApJ__MHD_SN,Bisnovatyi-Kogan_Popov_Samokhin__1976__APSS__MHD_SN,Mueller_Hillebrandt__1979__AA__MHD_SN}. Very
likely, the foremost effect that the combined action of magnetic
fields and rotation may bring is the exponential amplification of seed
fields mediated by the magneto-rotational instability (MRI).  The MRI
is the most promising candidate among the different possibilities for
field amplification. Other alternatives, \eg amplification of the
magnetic field due to the core compression, winding of the poloidal
into toroidal field, or hydrodynamic instabilities (convection and the
standing accretion shock instability, SASI), are less effective in
producing dynamically relevant magnetic fields after core bounce
\cite{Obergaulinger_et_al__2014__mnras__Magneticfieldamplificationandmagneticallysupportedexplosionsofcollapsingnon-rotatingstellarcores}.
The possibility that the MRI indeed was operative in the context of core
collapse opened up in simplified, spherically symmetric models
\cite{Akiyama_etal__2003__ApJ__MRI_SN}, in which large regions of the
post-bounce core were identified as MRI unstable. That seminal work
spurred various studies of axisymmetric collapse examining this result
more closely and exploring the consequences of a strong field created
by the MRI.  Some of the earlier works had used very approximate
equations of state (EOSs) and neglected neutrinos altogether
\cite{Obergaulinger_Aloy_Mueller__2006__AA__MR_collapse,Obergaulinger_et_al__2006__AA__MR_collapse_TOV,Cerda-Duran_et_al__2007__AA__passive-MHD-collapse,Sawai_Kotake_Yamada__2008__apj__Equatorially_asym_MHD_SN},
others treated neutrinos in a simplified manner
\cite{Ardeljan_et_al__2004__Astrophysics__2d_rot_collapse_Lagrangian_Coo,Ardeljan_Bisnovatyi-Kogan_Moiseenko__2005__MNRAS__MHD_SN,Moiseenko_et_al__2006__mnras__A_MR_CC_model_with_jets,Kotake_etal__2004__Apj__SN-magrot-neutrino-emission,Suwa_etal__2007__pasj__Magnetorotational_Collapse_of_PopIII_Stars},
while a few studies were performed with state-of-the-art neutrino
transport
\cite{Burrows_etal__2007__ApJ__MHD-SN,Dessart_et_al__2007__apj__MagneticallyDrivenExplosionsofRapidlyRotatingWhiteDwarfsFollowingAccretion-InducedCollapse,Obergaulinger_Aloy__2017__mnras__Protomagnetarandblackholeformationinhigh-massstars}.

These simulations commonly show the development of jets driven by the
magnetic field and powered by the rotational energy of the core if the
energy density of the magnetic field is locally similar to that of the
velocity field.  Although this possibility cannot be excluded in
principle, the weak pre-collapse fields predicted by current stellar
evolution modelling
\cite{Heger_et_al__2005__apj__Presupernova_Evolution_of_Differentially_Rotating_Massive_Stars_Including_Magnetic_Fields}
make it exceedingly unlikely to reach such a configuration without the
presence of efficient field amplification after bounce.  Following the
evolution of the MRI is technically very challenging because the length
scales on which it grows fastest is directly proportional to the seed
field strength.  On the numerical grids commonly employed in core
collapse models, these scales can only be resolved if the initial
field is already very strong, in which case the need for additional
MRI-mediated amplification is less pronounced.  Consequently, such
simulations cannot address the question whether the MRI is able to
amplify a realistically weak pre-collapse field to dynamically
relevant strength.  Instead,
\cite{Obergaulinger_etal__2009__AA__Semi-global_MRI_CCSN,Masada_et_al__2012__apj__LocalSimulationsoftheMagnetorotationalInstabilityinCore-collapseSupernovae,Rembiasz_et_al__2016__mnras__Terminationofthemagnetorotationalinstabilityviaparasiticinstabilitiesincore-collapsesupernovae,Rembiasz_et_al__2016__mnras__Onthemaximummagneticfieldamplificationbythemagnetorotationalinstabilityincore-collapsesupernovae}
performed local or semi-global three-dimensional simulations derived
from the shearing-box models of accretion-disk theory
\cite{Balbus_Hawley__1992__ApJ__MRI_3} with a reduced amount of
physics ingredients, but very high resolution.  The results, in
particular when combined with an analytical treatment of the MRI
growth, indicate that the factor by which the MRI can amplify the seed
field is limited by the development of secondary, parasitic
instabilities of Kelvin-Helmholtz type
\cite{Rembiasz_et_al__2016__mnras__Onthemaximummagneticfieldamplificationbythemagnetorotationalinstabilityincore-collapsesupernovae}.
More recently, growing computational power has allowed for an
increased resolution in global models specifically aimed at studying
the MRI, albeit with greatly simplified physics and only in
axisymmetry
\cite{Sawai_et_al__2013__apjl__GlobalSimulationsofMagnetorotationalInstabilityintheCollapsedCoreofaMassiveStar,Sawai_Yamada__2016__apj__TheEvolutionandImpactsofMagnetorotationalInstabilityinMagnetizedCore-collapseSupernovae}
or with a very restricted three-dimensional geometry
\cite{Masada_et_al__2015__apjl__MagnetohydrodynamicTurbulencePoweredbyMagnetorotationalInstabilityinNascentProtoneutronStars}.
Hence, the question of MRI-driven field amplification remains open.

Aside from leading to a wrong hydrodynamic turbulent cascade and
restricting the development of important non-axisymmetric flows such
as the spiral modes of the SASI, the two-dimensional, axisymmetric
setup of most simulations performed so far suppresses magnetic
dynamos, which may severely affect the field growth.  Hence, the move
to three-dimensional modelling, implying significant changes in the
dynamics of non-magnetized core collapse, is possibly even more
important in simulations with magnetic fields.  Strong growth of the
magnetic field due to three-dimensional SASI modes was found by
\cite{Endeve_et_al__2010__apj__Generation_of_Magnetic_Fields_By_the_SASI,Endeve_et_al__2012__apj__TurbulentMagneticFieldAmplificationfromSpiralSASIModes:ImplicationsforCore-collapseSupernovaeandProto-neutronStarMagnetization}
in a simplified setup.  Most recently,
\cite{Moesta_et_al__2015__nat__Alarge-scaledynamoandmagnetoturbulenceinrapidlyrotatingcore-collapsesupernovae}
obtained a magneto-rotational dynamo in a rapidly rotating core in
simulations with a leakage scheme with heating terms for the
neutrinos.  Using the same setup, they had previously already
shown another possibility of three-dimensional evolution, \viz the
instability of a magneto-rotational jet against kink modes
\cite{Mosta_et_al__2014__apjl__MagnetorotationalCore-collapseSupernovaeinThreeDimensions}.
In this aspect, their work extends the three-dimensional models of
\cite{Scheidegger_etal__2008__aap__GW_from_3d_MHD_SN,Winteler_et_al__2012__apjl__MagnetorotationallyDrivenSupernovaeastheOriginofEarlyGalaxyr-processElements}
that also had studied MHD jet formation in rotating core collapse.

For very much the same reasons as in the case of core collapse,
stellar evolution calculations, in particular when including magnetic
fields, should ideally
be carried out in multi-dimensional
geometry.
Though the very long evolutionary time scales
of stars in hydrostatic
equilibrium make such an approach very difficult in general, its
potential merits in the field
of
core collapse have been highlighted by
simulations of late burning stages
\cite{Mueller_et_al__2016__apj__TheLastMinutesofOxygenShellBurninginaMassiveStar,Cristini_et_al__2017__mnras__3Dhydrodynamicsimulationsofcarbonburninginmassivestars}
and the implications of a non-spherical structure of the core at the
onset of collapse for the explosion
\cite{Couch_Ott__2013__apjl__RevivaloftheStalledCore-collapseSupernovaShockTriggeredbyPrecollapseAsphericityintheProgenitorStar}.
With the equivalent of such multi-dimensional
studies for magnetized
stars not yet available and only few pre-collapse models with magnetic
fields in a spherically symmetric approximation existing, most studies
of core collapse with magnetic fields prescribe the pre-collapse
distributions of magnetic field and rotational velocity in terms of
simple functions with, \eg, parametrized decline with radius and
normalization.  These parameters are usually chosen such as to
approximate likely upper or lower bounds to the conditions in stellar
cores and to determine values for which the impact on the dynamics of
the core is relevant.

Here we follow this approach and simulate the evolution of the core of
a star with an initial mass of $20 \, \msol$ to which we added
different magnetic fields and rotational profiles.  We perform
axisymmetric simulations including special relativistic MHD, an
approximately relativistic gravitational potential, and a
sophisticated treatment of neutrino transport in order to characterize
the modifications induced by magneto-rotational effects on the
dynamics of the collapse.  Our particular focus lies on versions of
the basic model with rapid rotation and/or with strong magnetic
fields.  For those, we try to address questions such as:
\begin{itemize}
\item How do rotation and magnetic field produce explosions in a core
  that otherwise fails to explode?
\item How is neutrino heating
    affected? Does it play a major role in these cases?
\item How are the field and the angular momentum distributed across
  different layers of the \pns (PNS)?
\end{itemize}

In an attempt to enlighten the previous questions, this article is
organized as follows: we will introduce the physics and numerics of
our models as well as the initial conditions in \secref{Sek:PhysNum},
show the simulation results in \secref{Sek:Res}, and present a summary
and conclusions in \secref{Sek:SumCon}.

\section{Physical model, numerical methods and simulation setup}
\label{Sek:PhysNum}

Our models combine special relativistic magnetohydrodynamics (MHD), a
Newtonian gravitational potential with corrections approximating
general relativistic gravity, two-moment neutrino transport, and the
relevant reactions between neutrinos and matter.

We solve the equations of special relativistic MHD, i.e.~the
conservation laws for relativistic mass density, $D$, partial
densities of charged particles (electrons and protons), $Y_e D$, and
of a set of chemical elements, $X_k D$, relativistic momentum and
energy density, $\vec S$ and $\tau$, resp., and magnetic field,
$\vec B$, in the following form: 
\begin{eqnarray}
  \label{Gl:MHD-rho}
  \partial_{t} D 
  + \vec \nabla \alpha D \vec v 
  & = & 0,
  \\ 
  \label{Gl:MHD-Yrho}
  \partial_{t} Y_e D 
  + \vec \nabla \alpha Y_e D \vec v 
  & = & \alpha Q^{Y_e}_{\star},
  \\ 
  \label{Gl:MHD-Ycmp}
  \partial_{t} X_k D 
  + \vec \nabla \alpha X_k D \vec v 
  & = & R_k,
  \\ 
  \label{Gl:MHD-mom}
  \partial_{t} S^i
  + \nabla_j \alpha \mathcal{T}^{ij}
  & = & \alpha Q^{i}_{\star} - D \nabla^i \Phi, 
  \\ 
  \label{Gl:MHD-erg}
  \partial_{t} \tau
  + \vec \nabla \alpha \vec F_{\tau}
  & = & \alpha Q^{0}_{\star} + \alpha v_i Q^i_\star - S_i \nabla^i \Phi,
  \\ 
  \label{Gl:MHD-ind}
  \partial_{t} \vec B 
  + \vec \nabla \times \alpha ( \vec v \times \vec B ) 
  & = & 0,
  \\ 
  \label{Gl:MHD-divb}
  \vec \nabla \cdot \vec B  & =  & 0.
\end{eqnarray}
The operator $\nabla_i = \frac{1}{\sqrt{\gamma}}\partial_i
\sqrt{\gamma}$ contains the determinant of the spatial metric,
$\gamma$, which does not depend on time.  The relations between
conserved and \emph{primitive} variables are
\begin{eqnarray}
  \label{Gl:MHD-prim-rho}
  D & = & \rho W,
  \\
  \label{Gl:MHD-prim-S}
  S_i & = & ( \rho h + b^2 ) W^2 - b_i b^0,
  \\
  \label{Gl:MHD-prim-tau}
  \tau & = & ( \rho h + b^2 ) W^2 - ( P + b^2/2 ) - ( b^0)^2 - D,
\end{eqnarray}
where the primitive variables are the velocity, $\vec v$, the
rest-mass density, $\rho$, the specific enthalpy, $h = 1 + (
e_{\mathrm{int}} + P ) / \rho$, the internal energy density,
$e_{\mathrm{int}}$, and the gas pressure, $P$, and where the Lorentz
factor is defined as $W = (1 - v^2)^{-1/2}$; we use units in which the
speed of light is $c = 1$ in these previous equations.  Furthermore,
we have to take into account the relations holding for the magnetic
four-vector:
\begin{eqnarray}
  \label{Gl:MHD-prim-b0}
  b^0 & = & W B^i v_i,
  \\
  \label{Gl:MHD-prim-bi}
  b^i & = & ( B^i + b^0 W v^i ) / W.
\end{eqnarray}
We use the techniques for recovering the primitive variables described
in \cite{Cerda-Duran__2008__AA__GRMHD-code}.  Since the relations
inverting Eqs.\,(\ref{Gl:MHD-prim-rho})-(\ref{Gl:MHD-prim-tau}) are
not explicit, the momentum and energy fluxes are given in terms of a
combination of conserved and primitive variables:
\begin{eqnarray}
  \label{Gl:MHD-flux-S}
  \mathcal{T}^{ij} & = & 
  S^j v^i + \delta^{ij} ( P + b^2 / 2 ) - b^j B^i / W, 
  \\
  \label{Gl:MHD-flux-tau}
  F_{\tau}^i & = & 
  \tau v^i + ( P + b^2 / 2) v^i - b^0 B^i / W.
\end{eqnarray}
The other quantities appearing in the MHD equations are the lapse
function, $\alpha$, and the source terms accounting for the exchange
of lepton number, momentum, and energy between matter and neutrinos,
$Q^{Y_e}_{\star}, Q^{i}_{\star}$, and $Q^{0}_{\star}$, respectively.
The source terms are the integrals over neutrino energy, summed over
all neutrino flavours of the spectral neutrino-matter interaction
terms (see below).

The MHD system is closed by the SFHo equation of state of
\cite{Steiner_et_al__2013__apj__Core-collapseSupernovaEquationsofStateBasedonNeutronStarObservations}
for the gas above a density of $\rho = 6000 \, \gccm$.  At lower
densities, we use an equation of state containing contributions of
electrons, positrons, photons, and baryons as well as the flashing
scheme of \cite{Rampp_Janka__2002__AA__Vertex}.  The approximate way
in which this scheme changes the nuclear composition of the gas is
represented by the set of source terms $R_k$.

To compute the gravitational potential, $\Phi$, 
we solve the Poisson equation,
\begin{equation}
  \label{Gl:Poisson}
  \Delta \Phi_{\mathrm{N}} = 4 \pi G \rho
\end{equation}
for the two-dimensional Newtonian potential, $\Phi_{\mathrm{N}}$ and
replace its monopole component, $\Phi^{1d}_{\mathrm{N}}$ by the
approximately relativistic spherically symmetric potential,
$\Phi^{1d}_{\mathrm{TOV}}$:
\begin{equation}
  \label{Gl:Poisson-TOV}
  \Phi = \Phi_{\mathrm{N}} + ( \Phi^{1d}_{\mathrm{TOV}} - \Phi^{1d}_{\mathrm{N}}).
\end{equation}
For $\Phi^{1d}_{\mathrm{TOV}}$, we use version 'A' of the
post-Newtonian Tolman-Oppenheimer-Volkoff (TOV) potentials of
\cite{Marek_etal__2006__AA__TOV-potential}.  To ensure consistency
with the gravitational terms in the equations of neutrino transport
(see below), we include the lapse function in the spatial derivatives.
Since we do not model gravity by a general relativistic 3+1 metric, we
define the lapse function based on the classical gravitational
potential, $\Phi$, as $\alpha = \exp ( \Phi / c^2)$.  Our approach
represents a straightforward way to model the effects of relativistic
gravity in a non-GR code quite accurately.  We note that recently
genuinely general relativistic models of stellar core collapse have
become available (\eg
\cite{Kuroda_et_al__2012__apj__FullyGeneralRelativisticSimulationsofCore-collapseSupernovaewithanApproximateNeutrinoTransport,Ott__2013__apj__General-relativisticSimulationsofThree-dimensionalCore-collapseSupernovae,Kuroda_et_al__2016__apjs__ANewMulti-energyNeutrinoRadiation-HydrodynamicsCodeinFullGeneralRelativityandItsApplicationtotheGravitationalCollapseofMassiveStars,Mueller_et_al__2017__mnras__Supernovasimulationsfroma3Dprogenitormodel-Impactofperturbationsandevolutionofexplosionproperties}).

Neutrino transport is implemented in the two-moment framework
consisting of the conservation laws for neutrino energy and momentum
in the co-moving frame closed by the maximum-entropy Eddington factor
\cite{Cernohorsky_Bludman__1994__ApJ__MEC-Transport}.  We solve one
set of equations discretized in particle energy for each of the
three neutrino species (electron neutrinos, electron anti-neutrinos,
and heavy-lepton neutrinos).  The energy bins are coupled by velocity
terms and terms depending on the gravitational field, included in the
neutrino-transport equations in the $\mathcal{O}(v)$-plus formulation
of
\cite{Endeve_et_al__2012__ArXive-prints__ConservativeMomentEquationsforNeutrinoRadiationTransportwithLimitedRelativity}:
\begin{eqnarray}
  \label{Gl:neu-erg}
  \partial_{t} ( E +  v_i F^i )
  & + & \vec \nabla \alpha ( \vec F + v E )
  \\ 
  \nonumber
  & - & 
  ( \nabla_i \alpha + \dot{v}_i) \left[ \partial_{\epsilon} (\epsilon F^i) - F^i
  \right]
  \\ 
  \nonumber
  & - & 
  \nabla_i ( \alpha v_j )
  \left[ \partial_{\epsilon} ( \epsilon P^{ij}) - P^{ij}\right]
  \\ \nonumber
  & =  &
  \alpha Q_{0},
  \\
  %%%%%%%%
  \label{Gl:neu-mom}
  \partial_{t} ( F^i + v_j P^{ij} )
  & + & \nabla_j ( \alpha ( P^{ij} + v^j F^i) )
  \\ \nonumber
  & + & \dot{v}^i E + \alpha F^j \nabla_j v^i
  \\ \nonumber
  & + & ( E + P^j_j) \nabla^i \alpha
  \\ \nonumber
  & - & \partial_{\epsilon} (\epsilon P_{ij}) \dot{v}^j
  \\ \nonumber
  & - & \alpha \partial_{\epsilon} ( \epsilon U^{ki}_j) \nabla_k v^j
  \\ \nonumber
  & - & \partial_{\epsilon} ( \epsilon P^{ij}) \nabla_j \alpha
  \\ \nonumber
  & =  & 
  \alpha Q^i.
\end{eqnarray}
The momentum equation involves the third moment, $U$, for which we use
the approximation given in
\cite{Just_et_al__2015__mnras__Anewmultidimensionalenergy-dependenttwo-momenttransportcodeforneutrino-hydrodynamics},
where a thorough discussion of the implementation of the equations can
be found, too.

We include the following reactions between neutrinos and matter
(the implementation follows \cite{Rampp_Janka__2002__AA__Vertex}):
\begin{enumerate}
\item nucleonic absorption, emission, and scattering with the
  corrections due to weak magnetism and recoil \cite{Horowitz__2002__prd__Weakmagnetismforantineutrinosinsupernovae};
\item nuclear absorption, emission, and scattering;
\item inelastic scattering off electrons;
\item electron-positron pair annihilation into pairs of neutrinos and
  anti-neutrinos;
\item nucleon-nucleon bremsstrahlung.
\end{enumerate}
Recent studies
(\cite{Bollig_et_al__2017__PhysicalReviewLetters__MuonCreationinSupernovaMatterFacilitatesNeutrino-DrivenExplosions,Kotake_et_al__2018__apj__ImpactofNeutrinoOpacitiesonCore-collapseSupernovaSimulations,Burrows_et_al__2018__ssr__CrucialPhysicalDependenciesoftheCore-CollapseSupernovaMechanism})
highlighted the impact of using more sophisticated neutrino reaction
rates for the dynamics of core collapse, whose influence may go so far
as to induce explosions in (non-magnetized versions of) the same
progenitor we are using.

The flux terms of the equations of MHD and neutrino transport are of
hyperbolic character.  We therefore solve for them using standard
high-resolution shock-capturing methods with high-order spatial
reconstruction (monotonicity-preserving schemes,
\cite{Suresh_Huynh__1997__JCP__MP-schemes}) and approximate Riemann
solvers of HLL type.  To ensure a divergence-free evolution of the
magnetic field, we employ the upwind constrained transport method
\cite{Londrillo_Del_Zanna__2004__JCP__UpwindCT}.  Except for the
potentially stiff neutrino-source terms, the time integration is done using an
explicit $3^{\mathrm{rd}}$-order Runge-Kutta method.

The simulations were performed using a grid of $n_r = 480$ zones
logarithmically spaced in radial direction with a central grid width
of $\Delta r = 600 \, \mathrm{m}$ and a maximum radius of
$R_{\mathrm{out}} \approx \zehnh{9}{5} \, \km$ and
$n_{\mathrm{\theta}} = 128$ zones in angular direction.

The initial conditions of our simulations are based on the
pre-collapse model of a star of $20 \, \msol$ and solar metallicity
computed by
\cite{Woosley_Heger__2007__physrep__Nucleosynthesisandremnantsinmassivestarsofsolarmetallicity}.
This model, which is the result of one-dimensional hydrostatic stellar
evolution, includes neither rotation nor magnetic fields.  Thus, we
had to add both artificially when mapping the stellar evolution model
onto our axisymmetric simulation domain.  We consider three models:
\begin{description}
\item[\Modl{s20-1}] has a weak, non-vanishing magnetic field and a slow
  random angular velocity corresponding to a maximum value of
  $\Omega_{\mathrm{max; bounce}} \sim 0.1 \, \sek^{-1}$ at bounce. 
\item[\Modl{s20-2}] has the same magnetic field, but rotates at a
  moderate rate, corresponding to a maximum angular velocity at bounce
  of $\Omega_{\mathrm{max; bounce}} \sim 200 \, \sek^{-1}$.
\item[\Modl{s20-3}] rotates ten times faster than model \modl{s20-2}
  ($\Omega_{\mathrm{max; bounce}} \sim 2000 \, \sek^{-1}$) and
  possesses a stronger poloidal field than the latter, but the
  toroidal component was unchanged \wrt \modl{s20-2}.
\end{description}

We chose a cylindrical
rotational profile in which the angular frequency, $\Omega$, is a
function of the distance from the rotational axis, $\varpi$, only:
\begin{equation}
  \label{Gl:Init-Omega}
  \Omega = \Omega_0  \frac{ \varpi_{\Omega}^{q}} { \varpi^{q} + \varpi_{\Omega}^{q}}.
\end{equation}
We fix the exponent $q = 2$ and the cylindrical radius
$\varpi_{\Omega} = 10^8 \, \cm$ for all models and vary the
normalization of the angular velocity (see \tabref{Tab:init}).  The
magnetic field is initialized in a similar manner using a combination
of a poloidal component, $B^{\mathrm{P}}_0$, given in terms of the toroidal component of a
vector potential, $A^{\phi}$, and a toroidal component, $B^{\phi}_0$:
\begin{eqnarray}
  \label{Gl:Init-bpol}
  A^{\phi} & = & A_0^{\mathrm{{P}}} \frac{R_0^3}{R_0^3 + r^3} r \cos\theta,
  \\ 
  \label{Gl:Init-btor}
  b^{\phi} & = & B_0^{\phi} \frac{R_0^3}{R_0^3 + r^3} r \cos\theta
  .
\end{eqnarray}
We determine the normalization $A^{\mathrm{P}}_0$ such that the maximum
poloidal field has a given value $B^{\mathrm{P}}_0$.  The parameters of the
models are summarized in \tabref{Tab:init}, except for the one common
to all models, \viz $R_0 = \zehnh{2}{8} \, \cm$.

We note that the rotational frequencies of the latter two models are
relatively large when compared to typical results of stellar evolution
modelling with asymptotic (\ie, for $\varpi \gg \varpi_{\Omega}$)
specific angular momenta of $j_0^{\modelname{s20-1;s20-2}} =
\zehn{15;16} \, \cm^2 \, \sec^{-1}$ in the equatorial plane.  The
magnetic field strengths, too, are rather on the higher side of what
can be expected, albeit not tremendously enhanced.

\begin{table}
  \centering
  \begin{tabular}{l|ccc}
    \hline
    name & $\Omega_0 [\mathrm{Hz}]$ 
    & $\log B^{\mathrm{p}}_0 [\mathrm{G}]$ 
    & $\log B_0^{\phi} [\mathrm{G}]$
    \\
    \hline
    \modelname{s20-1} & random & $10$ & $11$
    \\
    \modelname{s20-2} & $0.1$ & $10$ & $11$
    \\
    \modelname{s20-3} & $1$ & $11$& $11$ 
    \\
    \hline
  \end{tabular}
  \caption{List of initial models and their initial parameters.  The columns
    display the model name and the normalizations of the angular velocity,
    the poloidal, and the toroidal components of the magnetic field. 
  }
  \label{Tab:init}
\end{table}

\section{Results}
\label{Sek:Res}

\begin{figure*}
  \centering
  \includegraphics[width=0.48\linewidth]{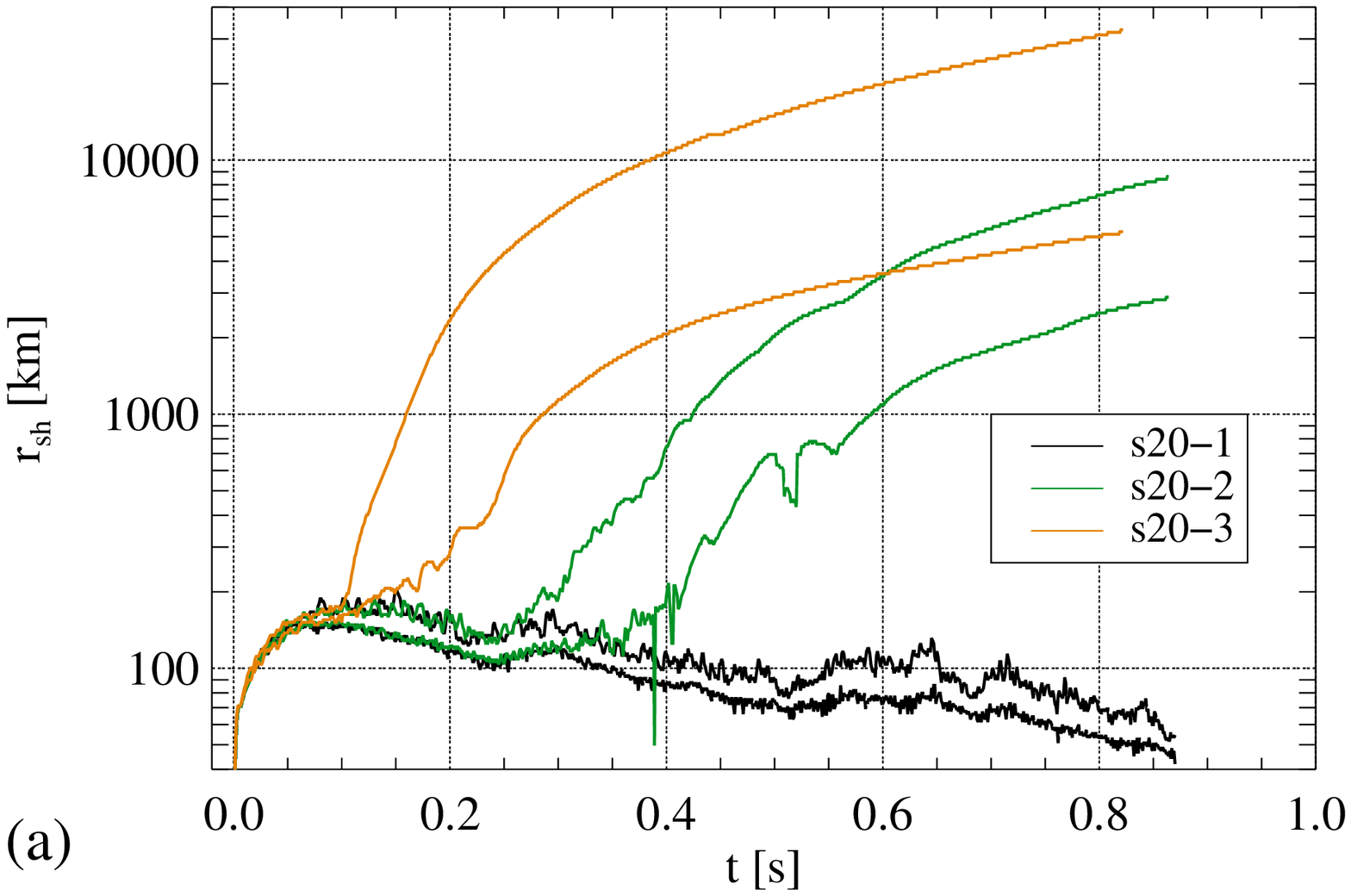}
  \includegraphics[width=0.48\linewidth]{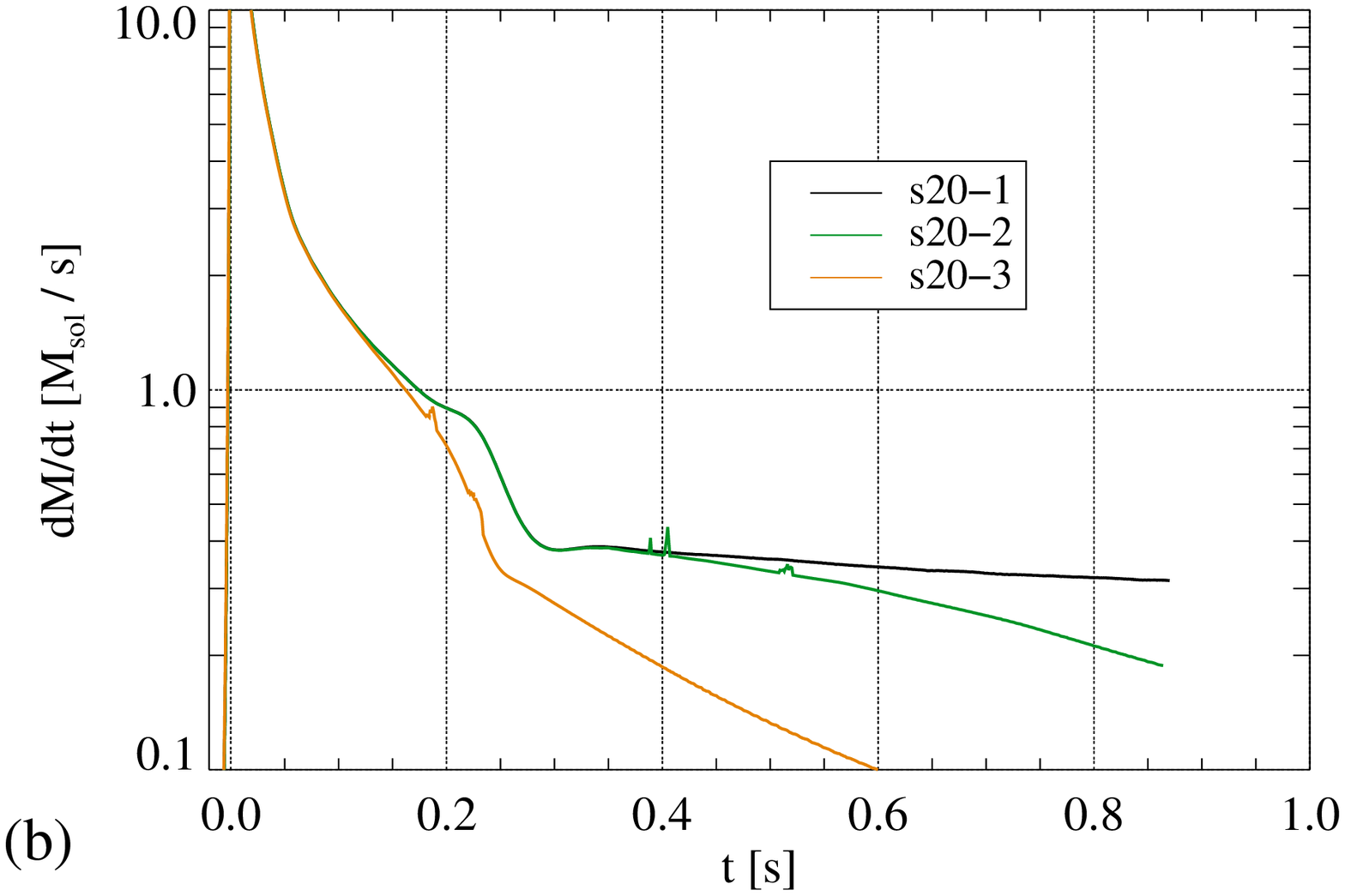}
  \includegraphics[width=0.48\linewidth]{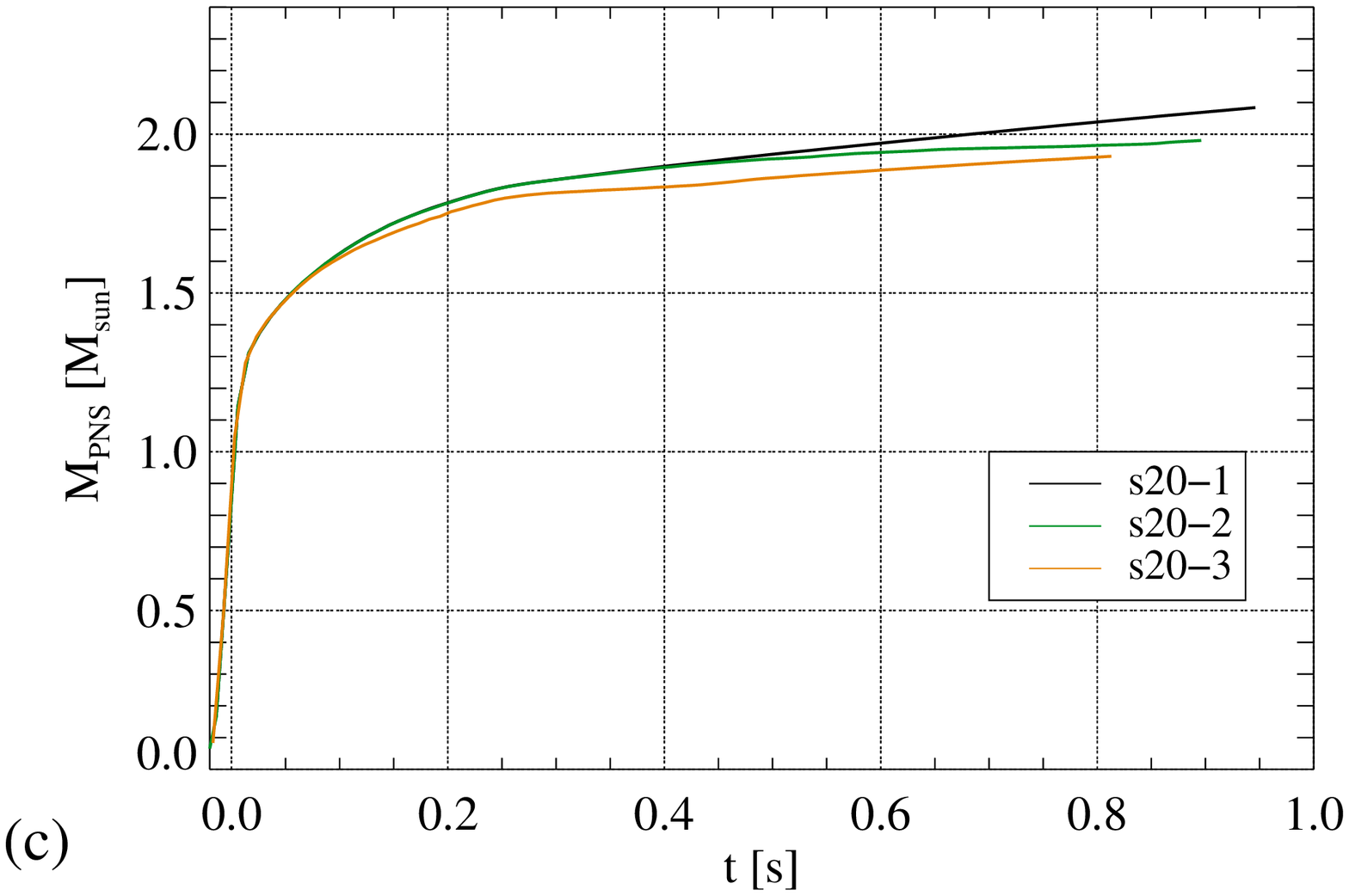}
  \includegraphics[width=0.48\linewidth]{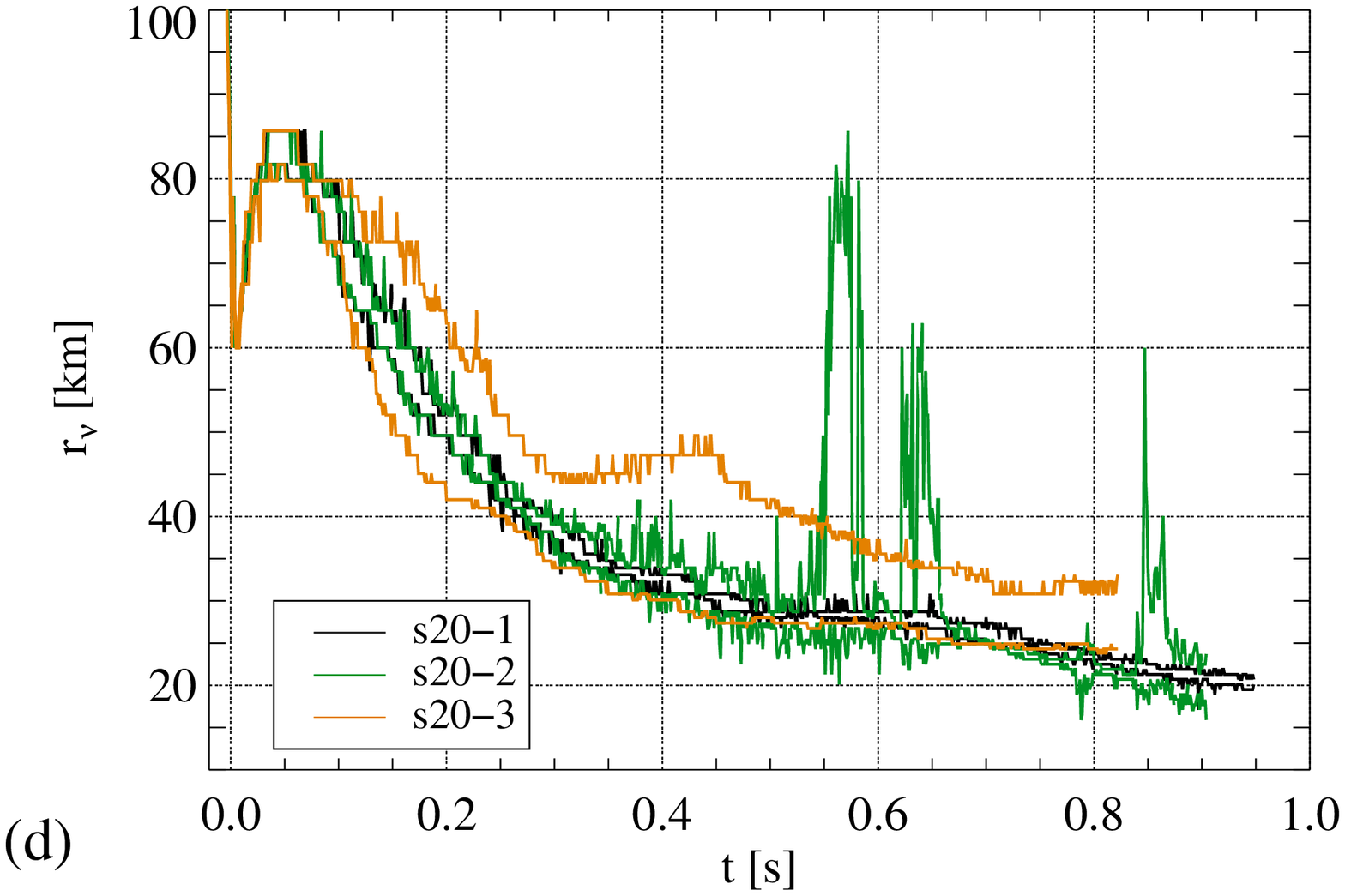}
  \caption{%%
    Comparison of quantities characterizing the three models,
    differentiated by colour, as functions of time.
    Panel (a): 
    minimum and maximum shock radii as a function of time.
    Panel (b): 
    mass accretion rate, measured at the shock radius.
    Panel (c): 
    mass of the PNS as a function of time.
    Panel (d): 
    minimum and maximum radii of the electron-\nusp.
  }
  \label{Fig:global}
\end{figure*}

\begin{figure*}
  \centering
  \includegraphics[width=0.48\linewidth]{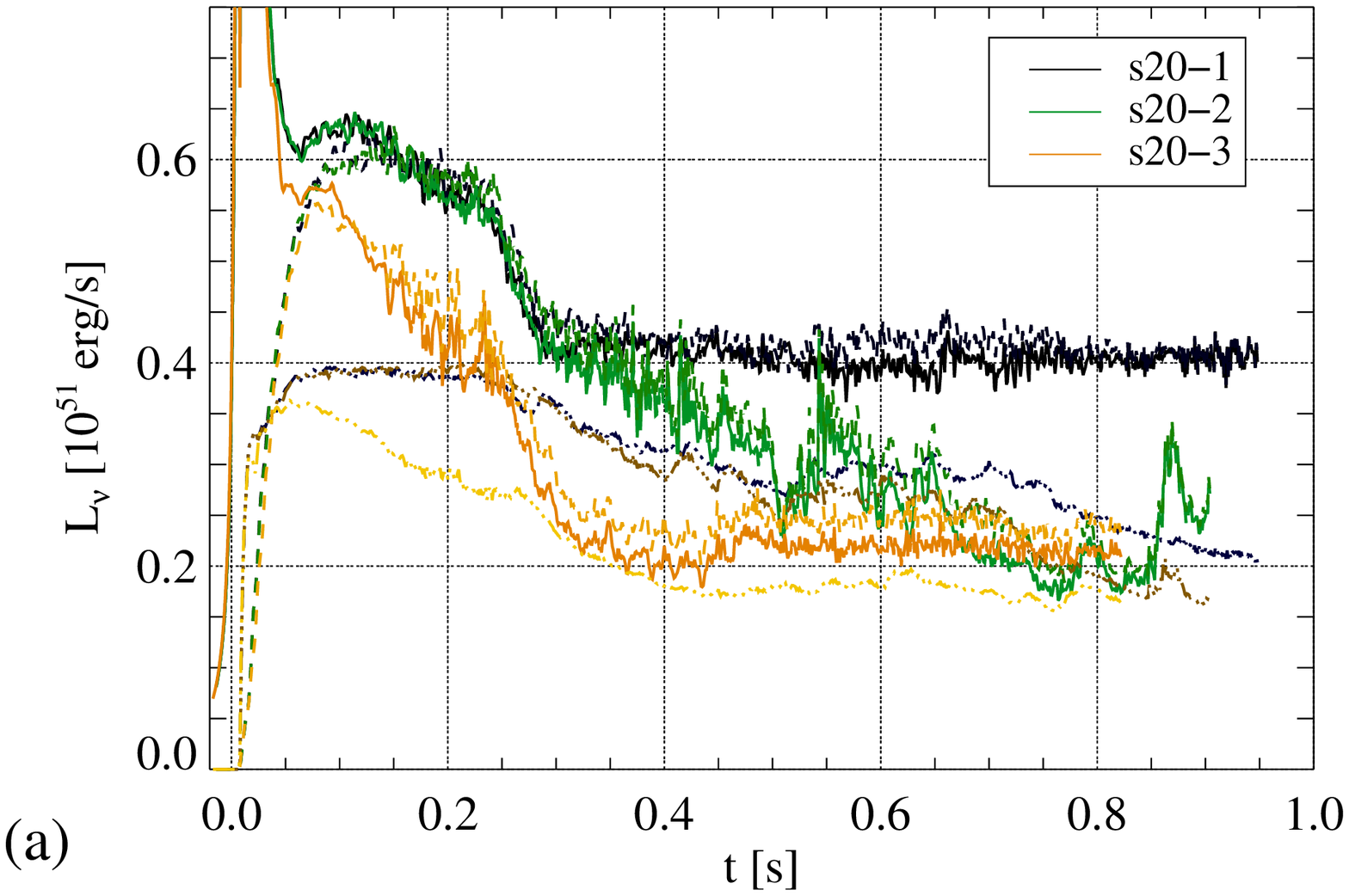}
  \includegraphics[width=0.48\linewidth]{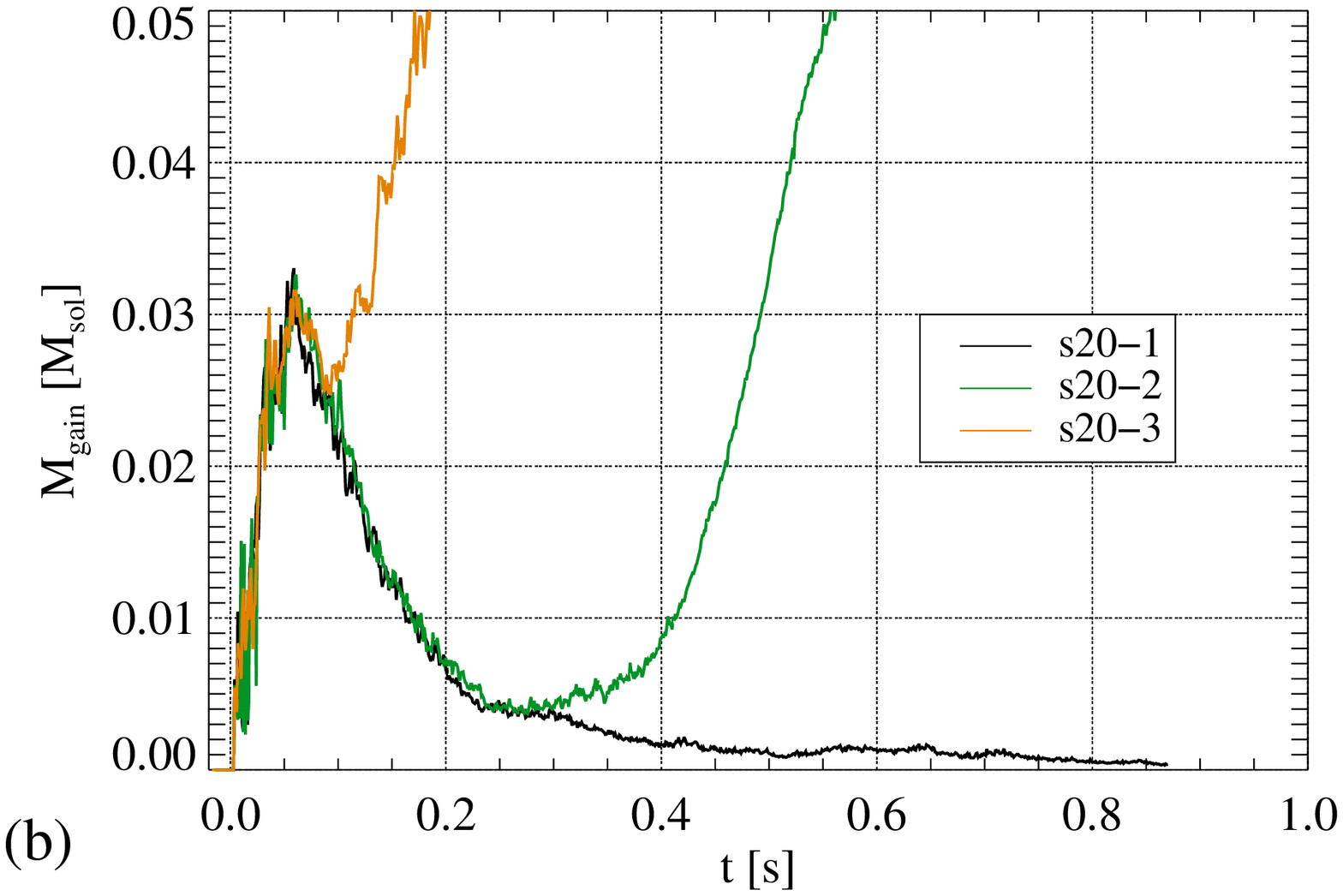}
  \includegraphics[width=0.48\linewidth]{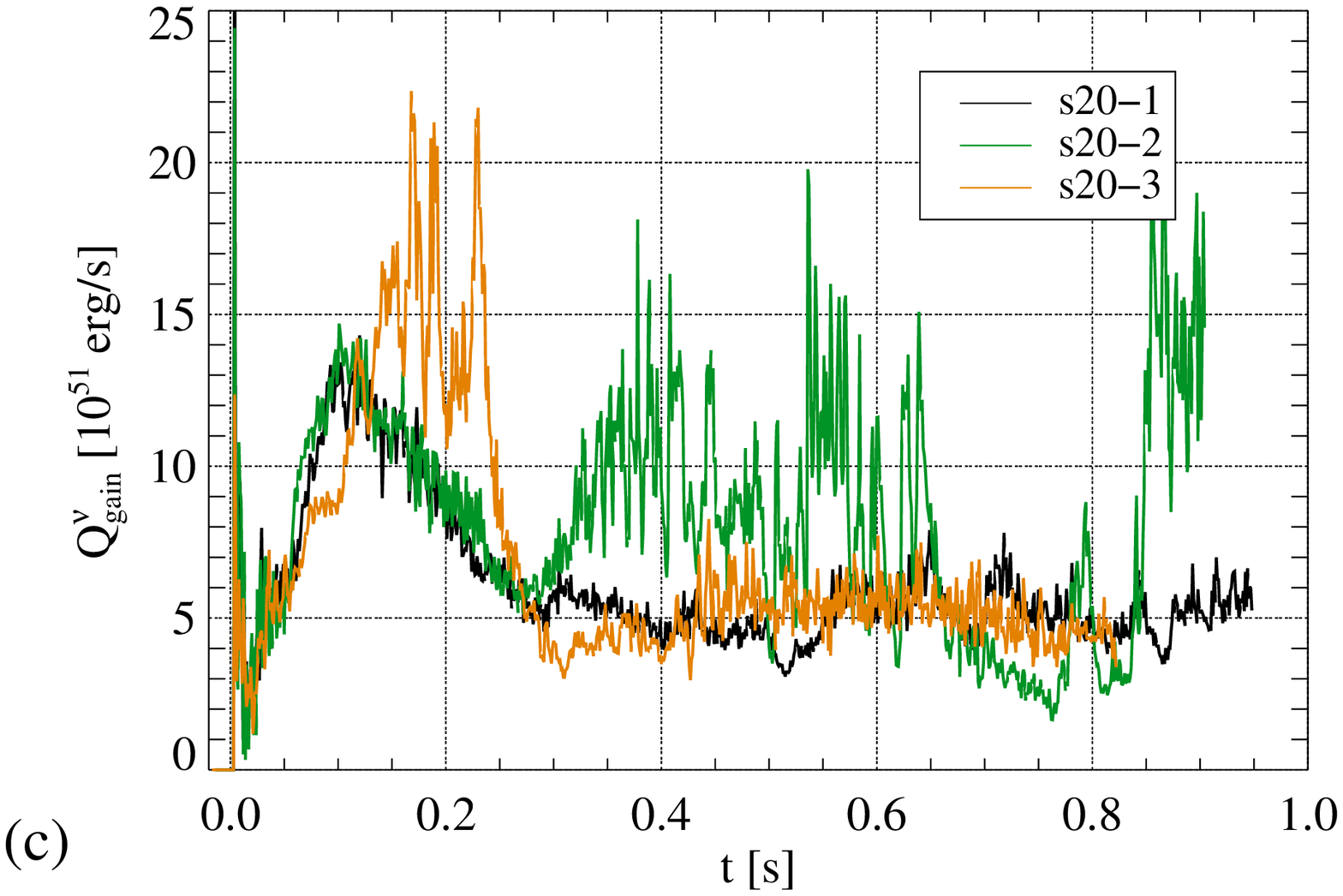}
  \includegraphics[width=0.48\linewidth]{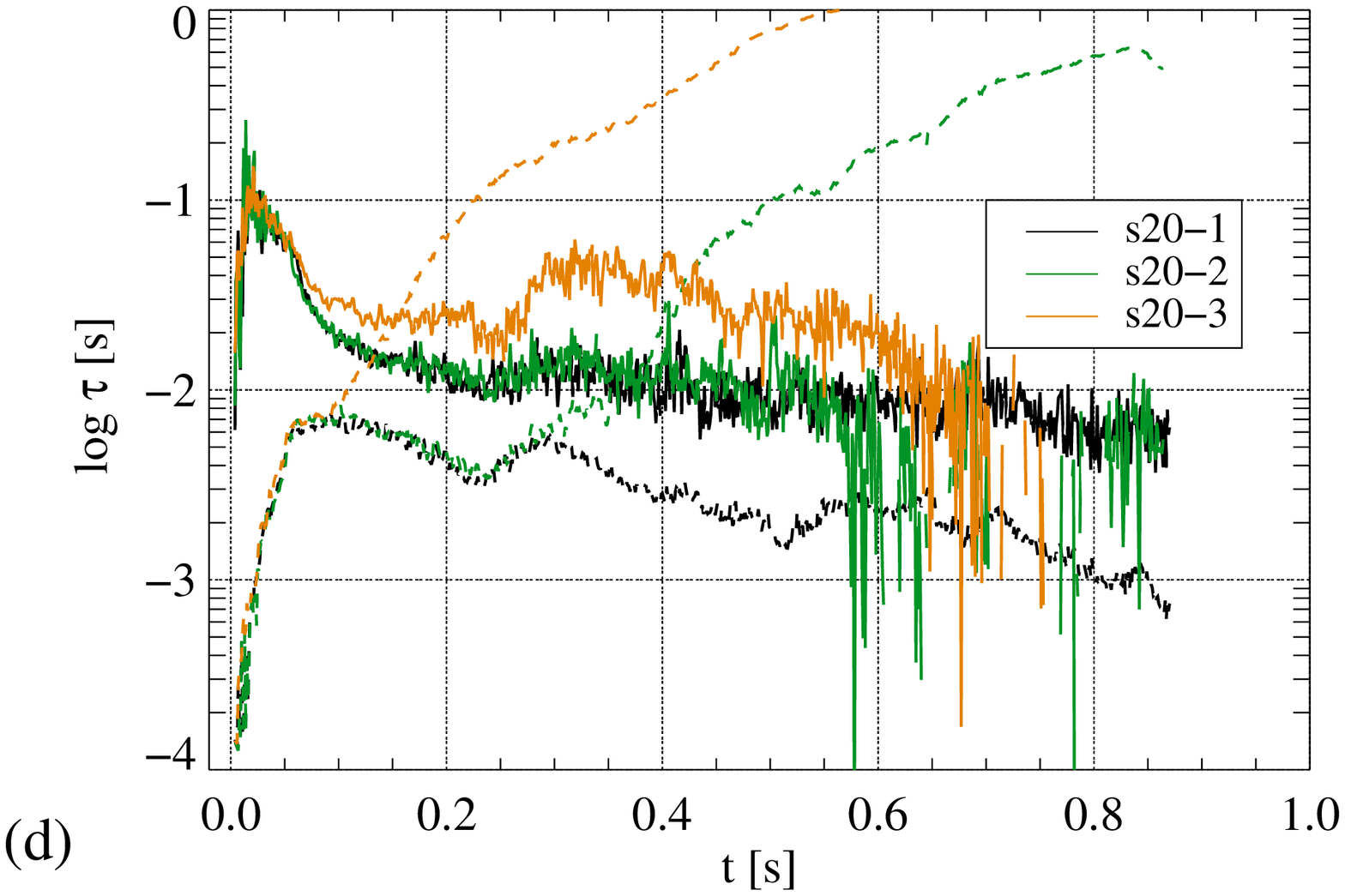}
  \includegraphics[width=0.48\linewidth]{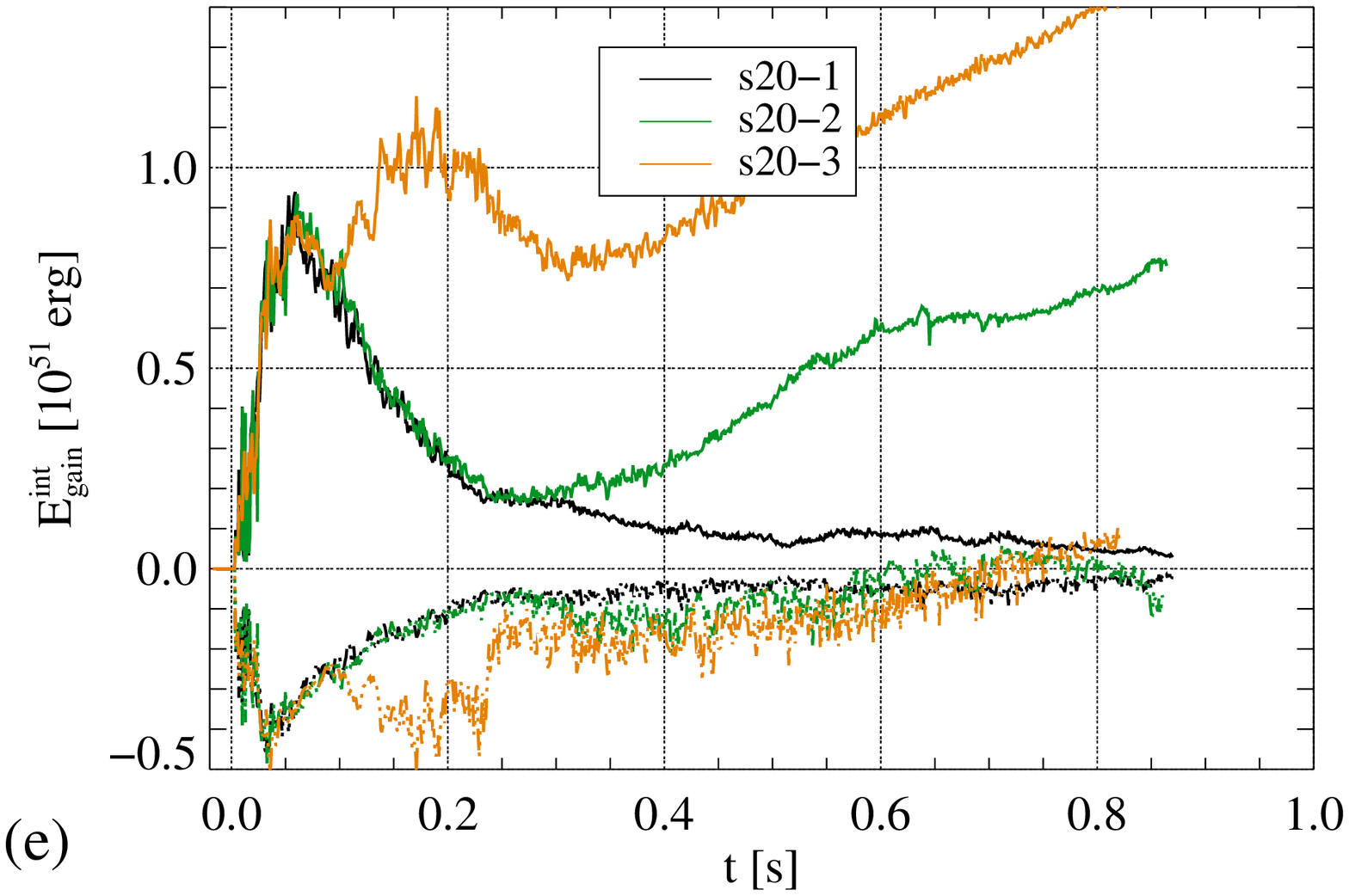}
  \includegraphics[width=0.48\linewidth]{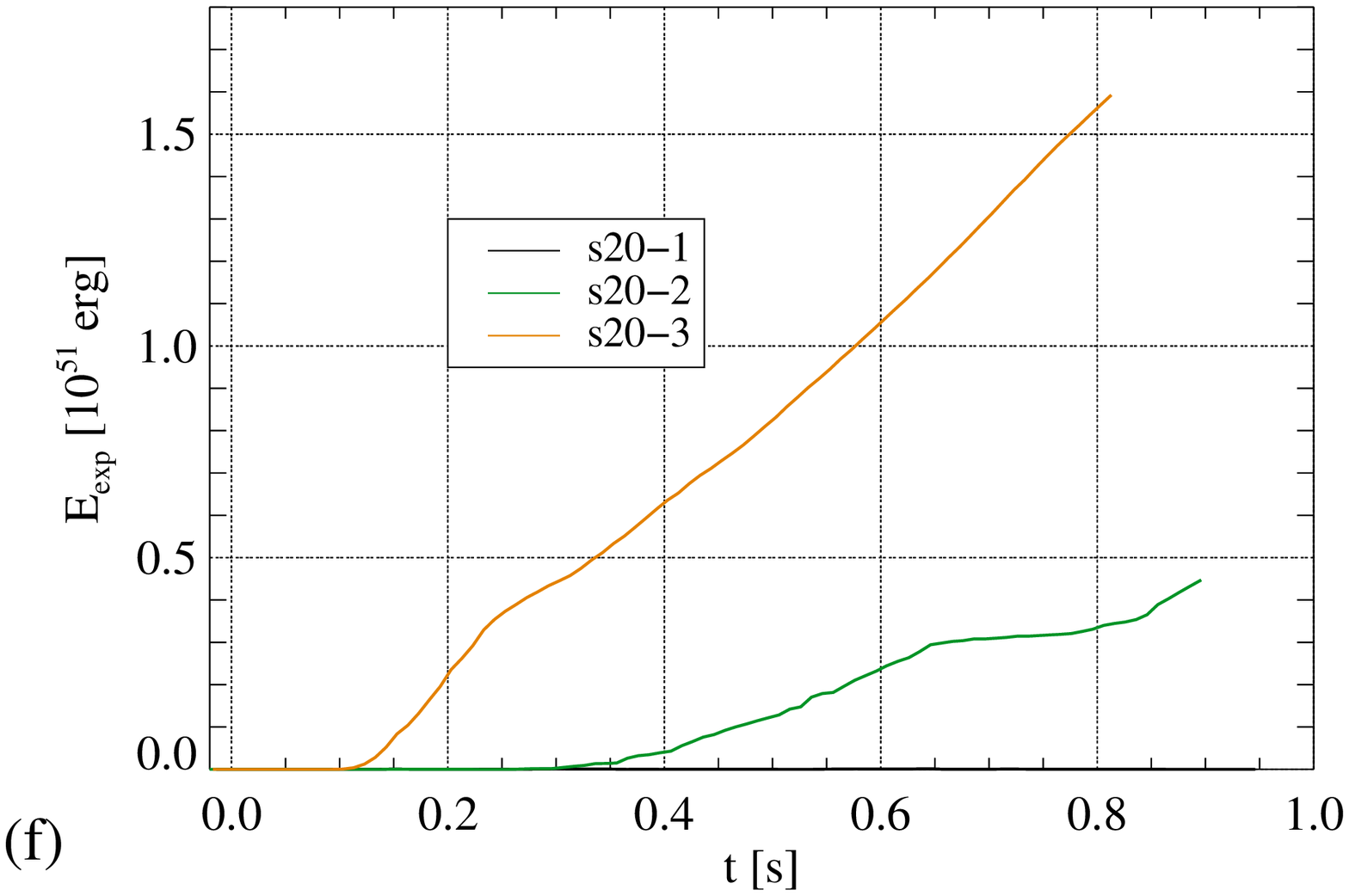}
  \caption{%%
    Continuation of \figref{Fig:global}.
    Panel (a): 
    lab-frame neutrino luminosities measured at a radius of $R_{\mathrm{Lum}} =
    500 \, \km$;  solid, dashed, and dash-triple-dotted lines
    correspond to $\nu_e$, $\bar{\nu}_e$, and $\nu_X$, respectively. 
    Panel (b): 
    mass contained in the gain layer.
    Panel (c): 
    heating rate in the gain layer.
    Panel (d): 
    advection (dash-triple-dotted lines) and heating (solid lines) time scales.
    Panel (e):
    internal (solid lines) and total (dashed-triple-dotted lines) energies in the gain layer.
    Panel (f):
    diagnostic explosion energy as a function of time (no line is
    visible for the non-exploding \modl{20-1}).
  }
  \label{Fig:global-2}
\end{figure*}

We present a number of global variables characterizing the time
evolution of the three models in Figs.~\ref{Fig:global} and
\ref{Fig:global-2}.  They serve us to explore the differences between
models, which will be discussed individually in the following.

\subsection{\Modl{s20-1}}
\label{sSek:res:s20-1}

The model for which the influence of rotation and magnetic fields is
negligible, \modl{s20-1}, does not explode.  As shown in \panel{a} of
\figref{Fig:global}, the shock wave stalls after $t \approx 100 \,
\msec$.  Afterwards, it starts to recede slowly, with only a brief
episode ($t \approx 240 ... 300 \, \msec$) in which the shock
contraction is interrupted, caused by the lower mass accretion rate
(see \panel{b}) onto the PNS as the Si interface of the core is
falling through the shock.  Owing to the accretion, the PNS mass grows
steadily and reaches $M_{\mathrm{PNS}} \approx 2.11 \, \msol$ at the
end of the simulation (panel \banel{(c)}).  The PNS might be stable
against its own self-gravity for several more seconds, but unless an
explosion develops will probably collapse to a black hole.

Throughout the entire evolution, the shock is moderately asymmetric
with a ratio between maximum and minimum radii fluctuating in the
range $6 : 5 \lesssim r_{\mathrm{sh;max}} : r_{\mathrm{sh;min}}
\lesssim 8:5$.  This asymmetry bears a clear imprint of north-south
sloshing modes exciting in the gain layer, with the location of the
maximum radius quasi-periodically oscillating from one pole to the
other and the equatorial radius showing only small variations.

The surface of the PNS, for which we take the electron-\nusp
(\panel{d}) as a proxy, continuously contracts from a maximum radius
$r_{\nu} \gtrsim 80 \, \km$ to one of $r_{\nu} \gtrsim 20 \, \km$ by
the end of the simulation at $t = 960 \, \msec$.  Owing to the minor
degree of rotation, the \nusp is essentially spherical.

The luminosity (\figref{Fig:global-2}, \panel{a}) of all flavours
after the $\nu_e$-burst, which lasts roughly until $t \approx 50 \,
\msec$, remains relatively high until the accretion of the interface
of the inner core around $t \approx 250 \, \msec$.  After that point,
the lower accretion rate translates into a lower release of
gravitational energy in the form of neutrinos.  $\nu_e$ and
$\bar{\nu}_e$ are emitted at constant (and almost equal) luminosities,
while the luminosity of the heavy-lepton neutrinos gradually
decreases.

To further understand the failure of the explosion, we turn our
attention to neutrino heating in the gain layer.  Along with the
contraction of the shock wave, the mass contained in the gain layer
(\panel{b} of \figref{Fig:global-2}) decreases from a maximum of
$M_{\mathrm{gain}} \approx 0.03 \, \msol$ attained at $t \approx 60 \,
\msec$ to below $M_{\mathrm{gain}} \lesssim 0.001 \, \msol$ at the end
of the simulation.  Except for an initial rising phase, the net
neutrino heating integrated over the gain layer (\panel{c}), parallels
the evolution of the neutrino luminosity and levels off at a value of
$Q^{\nu}_{\mathrm{gain}} \approx \zehnh{5}{51} \, \erg \sec^{-1}$.
Hence, for most of the evolution, the model combines a constant mass
accretion rate and a constant neutrino heating rate in a contracting
gain layer.  A consequence of this evolution is that the
gain-layer average of the neutrino heating time scale, defined in
terms of the internal and gravitational energy of the gain layer,
$E^{\mathrm{int/grav}}_{\mathrm{gain}}$, as
\begin{equation}
  \label{Gl:tauhtg}
  \tau_{\mathrm{htg}} = 
  \frac{E^{\mathrm{int}}_{\mathrm{gain}}  + E^{\mathrm{grav}}_{\mathrm{gain}}}  {Q^{\nu}_{\mathrm{gain}}}
  ,
\end{equation}
does not significantly fall below
$\tau_{\mathrm{htg}} \approx 10 \, \msec$ (see \figref{Fig:global-2},
\panel{d}).  The dwell time, $\tau_\mathrm{adv}$, \ie, the
value of the integral
$\tau_{\mathrm{adv}} (\theta) = \int_{\mathrm{gain \ layer}}
\mathrm{d} r / v^{r}$ averaged over polar angles, is shorter at all
times.  It increases to a first maximum at $t \approx 120 \, \msec$
and then decreases as the shock radii decrease faster than the gain
radii.  The subsequent slight expansion of the shock wave increases
the width of the gain layer.  The corresponding growth of
$\tau_{\mathrm{adv}}$, however, is not enough to reach the heating
time scale, and no explosion is launched.  After
$t \approx 300 \, \msec$, the shock contraction sets in again, and
$\tau_{\mathrm{adv}}$ decreases.  Even though it eventually rises
again towards the end of the simulation, it never exceeds
$\tau_{\mathrm{htg}}$.  The internal energy of the gain layer
(\figref{Fig:global-2}, \panel{e}) shows a steady decline after
$t \approx 60 \, \msec$, while its total energy increases from a
minimum obtained at the same time, but without ever attaining positive
values.  Thus, the gain layer is gravitationally bound at all times,
disfavoring shock revival.

\subsection{\Modl{s20-2}}
\label{sSek:res:s20-2}

\begin{figure}
  \centering
  \includegraphics[width=1.00\linewidth]{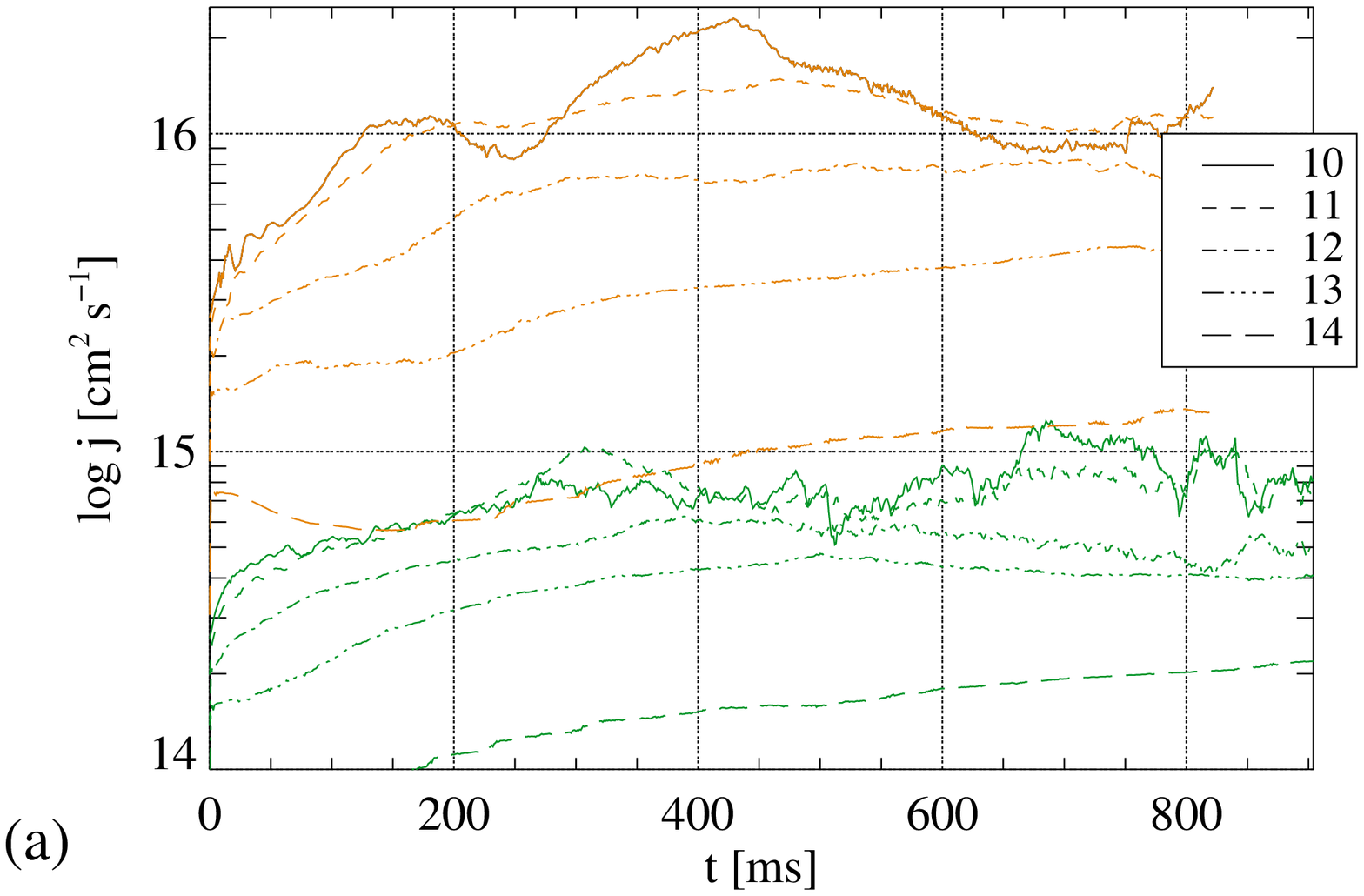}
  \includegraphics[width=1.00\linewidth]{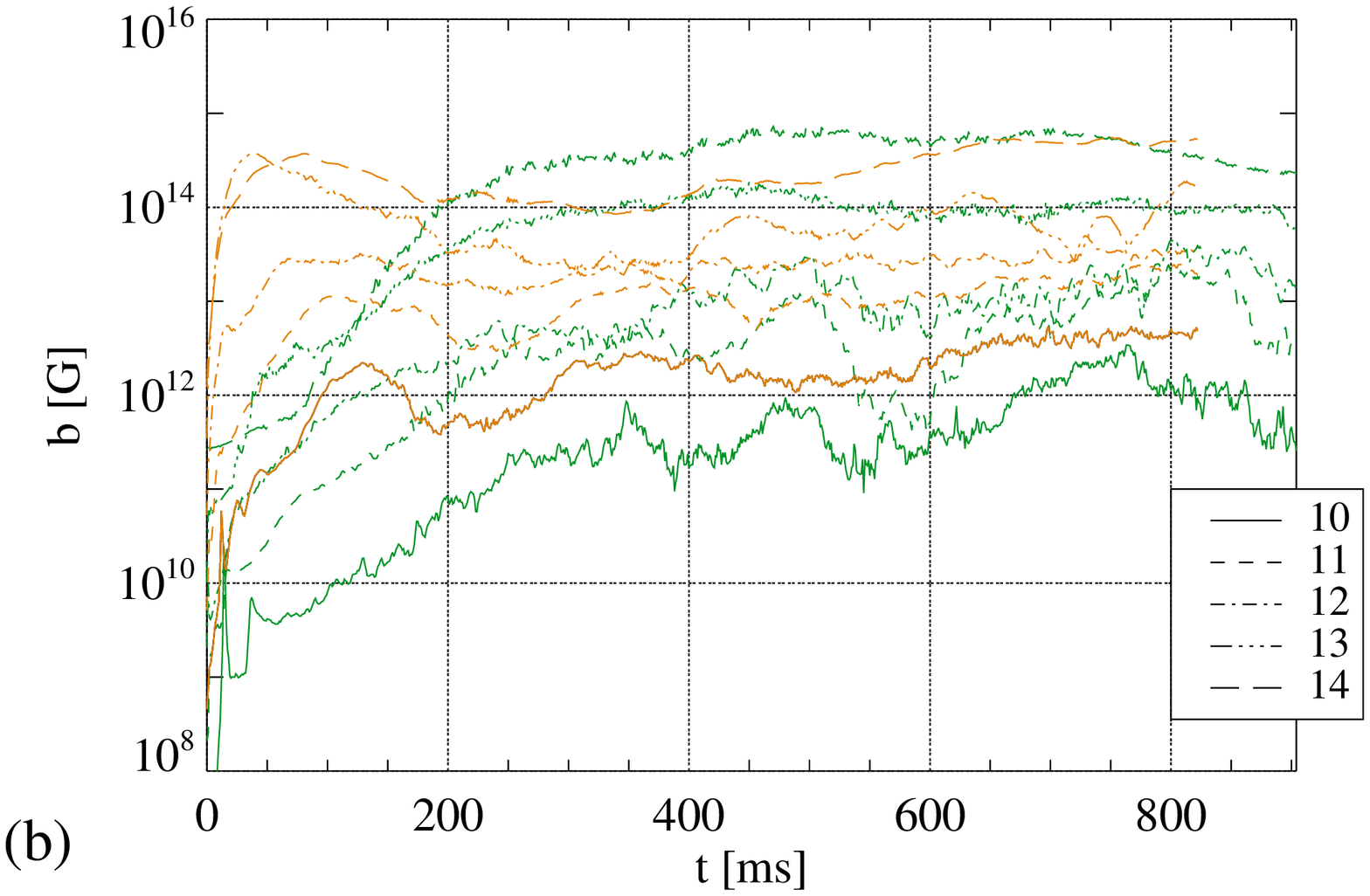}
  \includegraphics[width=1.00\linewidth]{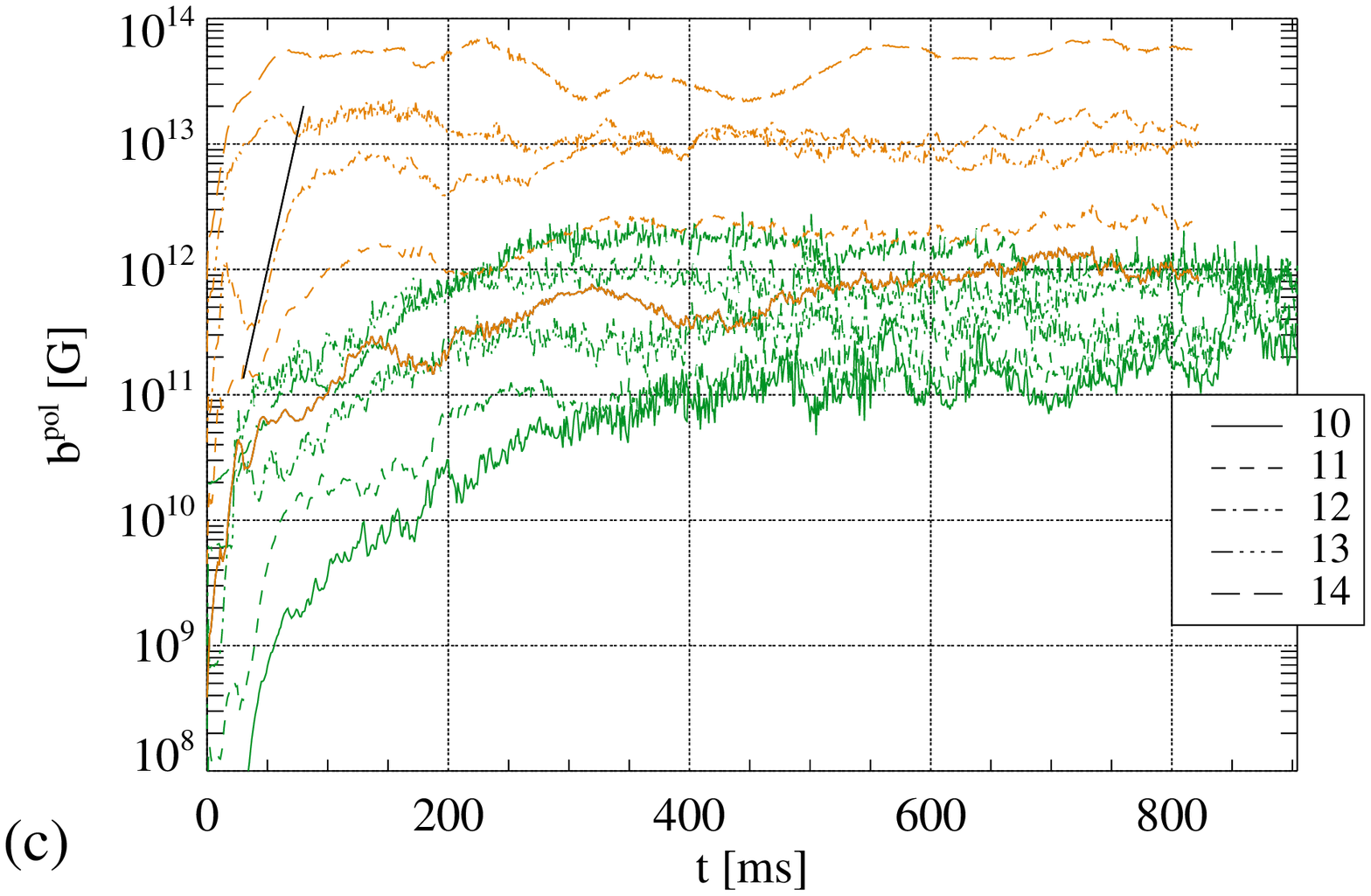}
  \caption{%%
    Panel \banel{(a)}: specific angular momentum of \modls{s20-2} (green
    lines) and \modelname{s20-3} (orange) as functions of time.
    Different lines correspond to averages over layers with densities
    $\log_{10}(\rho\, [\gccm]) \in [10, 14] \,
    $, as indicated in the legend.
    Panel \banel{(b)}: the average magnetic field strength as a function of
    time for the same layers of the two models.
    Panel \banel{(c)}: the same as before, but for the average poloidal
    magnetic field strength.  We add an
    exponential function (solid black line) to indicate the phase of
    MRI growth in the envelope of the PNS of \modl{s20-3}.
  }
  \label{Fig:s20-2--global}
\end{figure}

The most important differences of \modl{s20-2} from \modelname{s20-1}
are its fairly fast rotation and strong field.  We refer to \panel{a}
of \figref{Fig:s20-2--global} for the evolution of the specific
angular momentum contained in different layers of the PNS of this
model and \modl{s20-3}.  The temporal evolution of $j$ shows a growth
during the first epoch of the simulation and more or less constant
values afterwards.  From the innermost core with $\rho \ge 10^{14} \,
\gccm$, $j$ increases towards the PNS surface where it assumes a value
of $j \sim 10^{15} \, \cm^2 / \sec$, which is sufficient to alter the
dynamics of these regions.

The magnetic field, on the other hand, is strongest in the centre and
weakest at the surface (\panel{b} of \figref{Fig:s20-2--global}).
However, even there, $b$ reaches an average strength of $b \sim
10^{12} \, \Gauss$, dominated by the toroidal component with a
relatively weak poloidal component (bottom panel).  Locally, it is
even stronger which, as we shall see below, has important dynamical
consequences.

Early on, \modl{s20-2} evolves very similarly to \modl{s20-1}.  The
minimum and maximum radii of shock and electron-\nusp, the accretion
rate and the neutrino luminosities as well as the mass and neutrino
heating rate in the gain layer and the time scales of advection and
neutrino heating and the resulting energies show only stochastic
deviations from the model discussed above (see \figref{Fig:global} and
\ref{Fig:global-2}).

Unlike in \modl{s20-1}, the accretion of the first interface at $t
\approx 250 \, \msec$ triggers an explosion.  The expansion of the
shock radii setting in at $t \approx 220 \, \msec$ is sustained and,
rather than turning into another epoch of contraction, accelerates,
reaching a maximum radius of $r = 1000 \, \km$ about 200 ms after
shock revival.  The minimum shock radius increases in a similar
fashion as the maximum, albeit delayed by about 200 ms.  The explosion
geometry is clearly bipolar with an essentially symmetric pair of
outflows along the rotational axis and downflows at low latitudes.
With accretion occurring only under a limited range of angles, the
mass accretion rate gradually drops, reducing the neutrino
luminosities. Furthermore, the PNS mass grows slower than for
\modl{s20-1}.  At $t = 900 \, \msec$, it is still below $2 \, \msol$,
\ie, about $0.1 \msol$ less than in the former model.  The diagnostic
explosion energy, $E_{\mathrm{exp}}$, \ie the total (internal,
kinetic, magnetic plus gravitational) energy energy of gravitationally
unbound, regions (\figref{Fig:global-2}, panel \banel{(f)}) is
continuously rising after the onset of the evolution.  Hence, it is
still not converged by the end of the simulation and the final value
will exceed that of $E_{\mathrm{exp }} \approx \zehnh{4.6}{50} \,
\erg$ we find at $t \approx 900 \, \msec$.

The diagnostic quantities show behaviours consistent with the
development of an explosion.  Around the time of shock revival, the
neutrino heating rate in the gain layer increases.  While this
increase is not directly reflected in the heating time scale, we see a
growth of the dwell time until it exceeds the heating time.  The
internal energy of the gain layer increases and the total energy eventually
becomes positive.

To fully understand the mechanism leading to the explosion, we must,
however, go beyond these global variables and pay closer attention to
the details of the dynamics of the model.  Such an approach is
required because of the asymmetries of the core.  The fact that the
explosion starts at the poles while the equatorial regions are far
from shock runaway makes for a poor correlation between angularly
integrated quantities and the dynamics.  We see this, \eg, in the fact
that the equality between advection and heating times, is reached
around $t \approx 350 \, \msec$, \ie about 100 ms after the first
signs of the beginning of the explosion, \viz the initiation of the
shock expansion at $t \approx 250 \, \msec$.

\begin{figure*}
  \centering
  \includegraphics[width=0.48\linewidth]{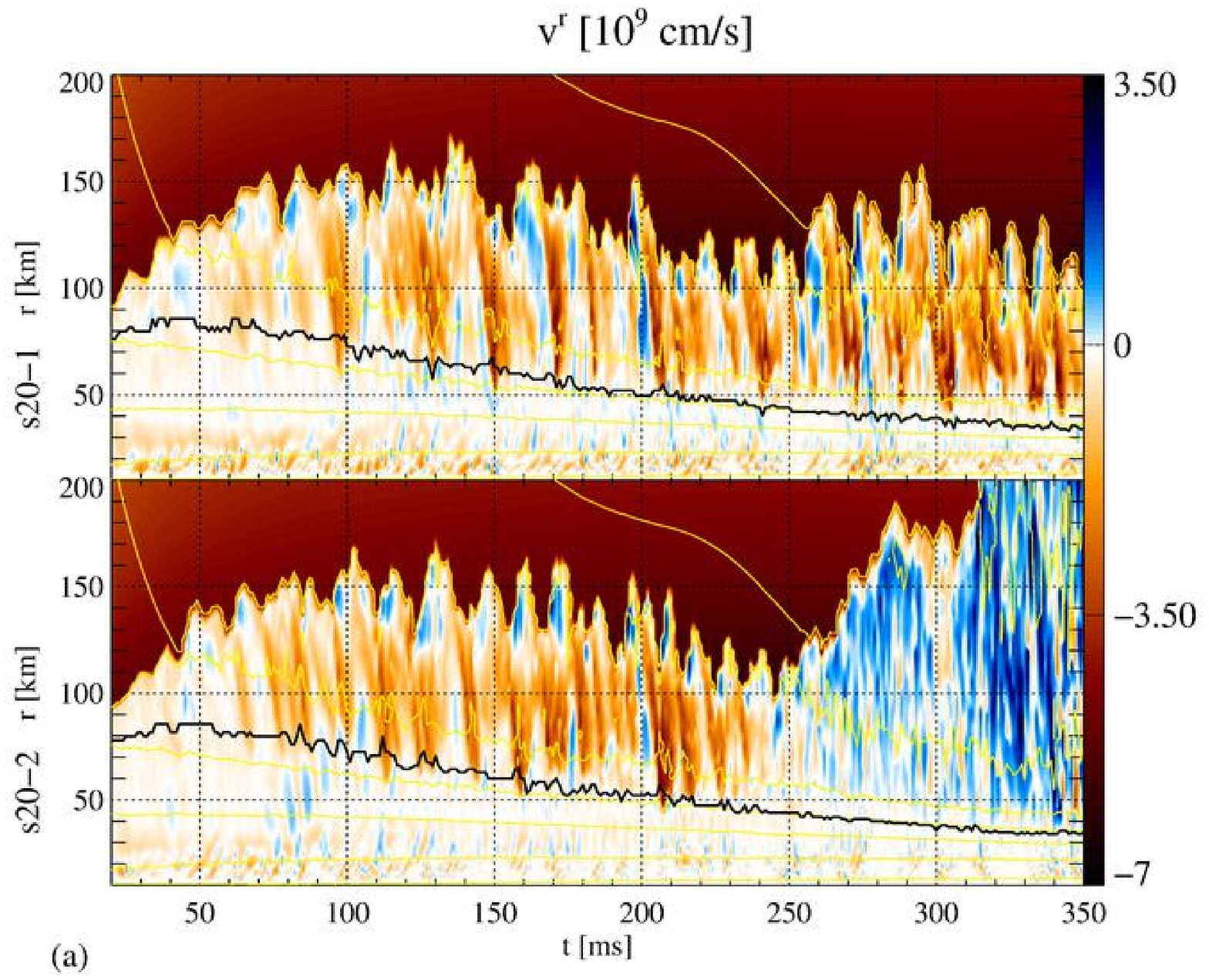}
  \includegraphics[width=0.48\linewidth]{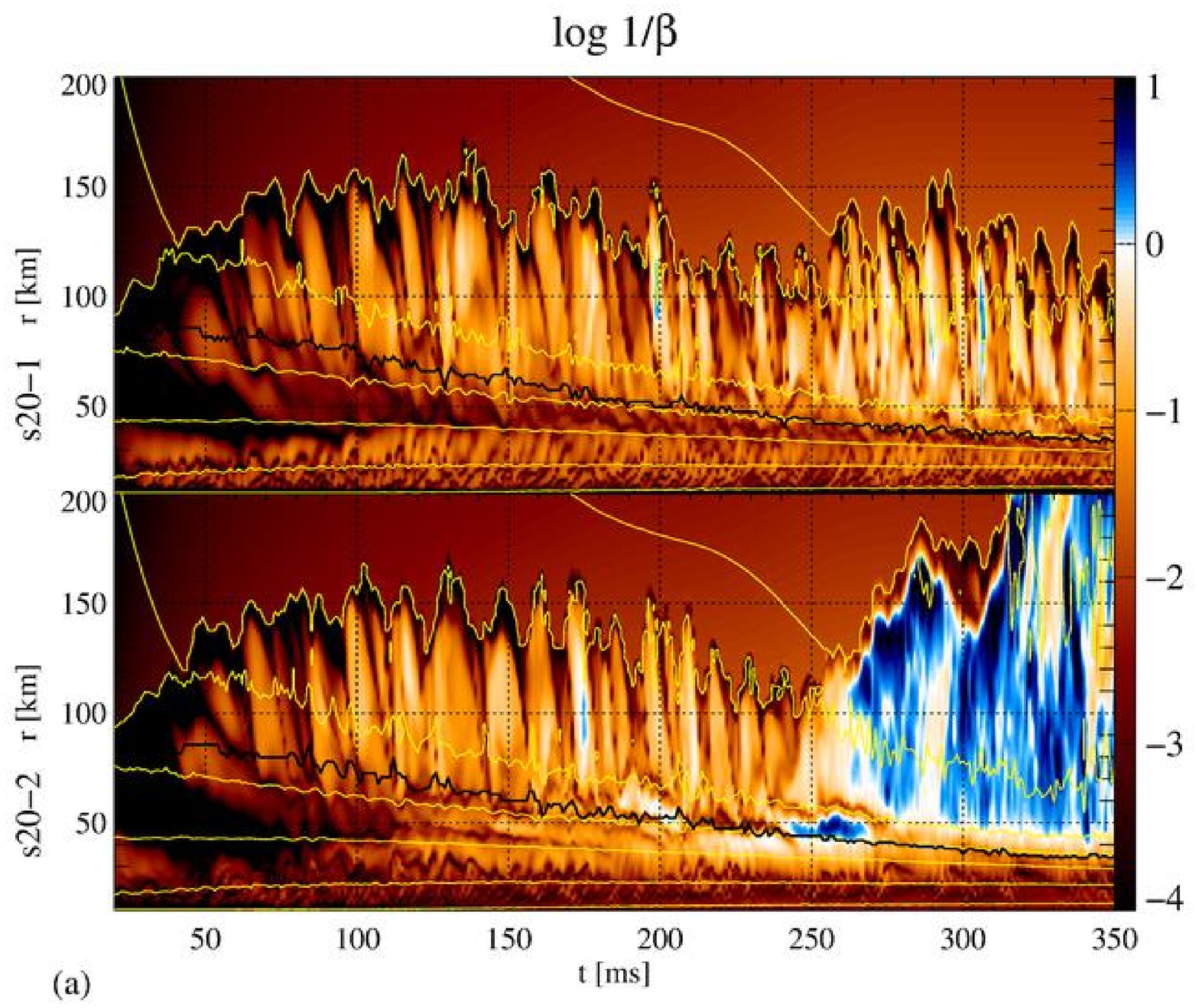}
  \caption{%%
    Evolution of the radial velocity (\panel{a}) and the
    inverse plasma-$\beta$ parameter (\panel{b}) along the south pole of
    \modls{s20-1} (upper parts of the panels) and \modelname{s20-2}
    (lower parts).  %%
    In both panels, the black lines represent the gain radius, while
    the yellow lines are iso-density contours.
}
  \label{Fig:s20-2--1}
\end{figure*}

Hence, we examine the dynamics of the south polar region where the
explosion is launched slightly earlier than at the north pole.  In
\figref{Fig:s20-2--1}, \panel{a}, we compare the evolution of the
radial velocity along the south pole at times $t \in [50, 300 ] \,
\msec$ after bounce for \modls{s20-1} (top part of the panel) and
\modelname{s20-2} (bottom half).  Until the drop in mass accretion
rate at $t \approx 250 \, \msec$, the post-shock region of both models
is characterized by downflows (reddish colours) with upflows (blue)
showing up only occasionally.  This pattern does not change
significantly after that time, \ie, in the brief period during which
the shock wave of \modl{s20-1} expands.  For \modl{s20-2}, on the
other hand, the onset of shock expansion is marked by a clear
predominance of positive radial velocities.  Their large-scale nature,
connecting the vicinity of the \nusp (black line) and the shock wave,
suggests to search for the mechanism driving shock revival near the
PNS.

\begin{figure*}
  \centering
  \includegraphics[width=0.49\linewidth]{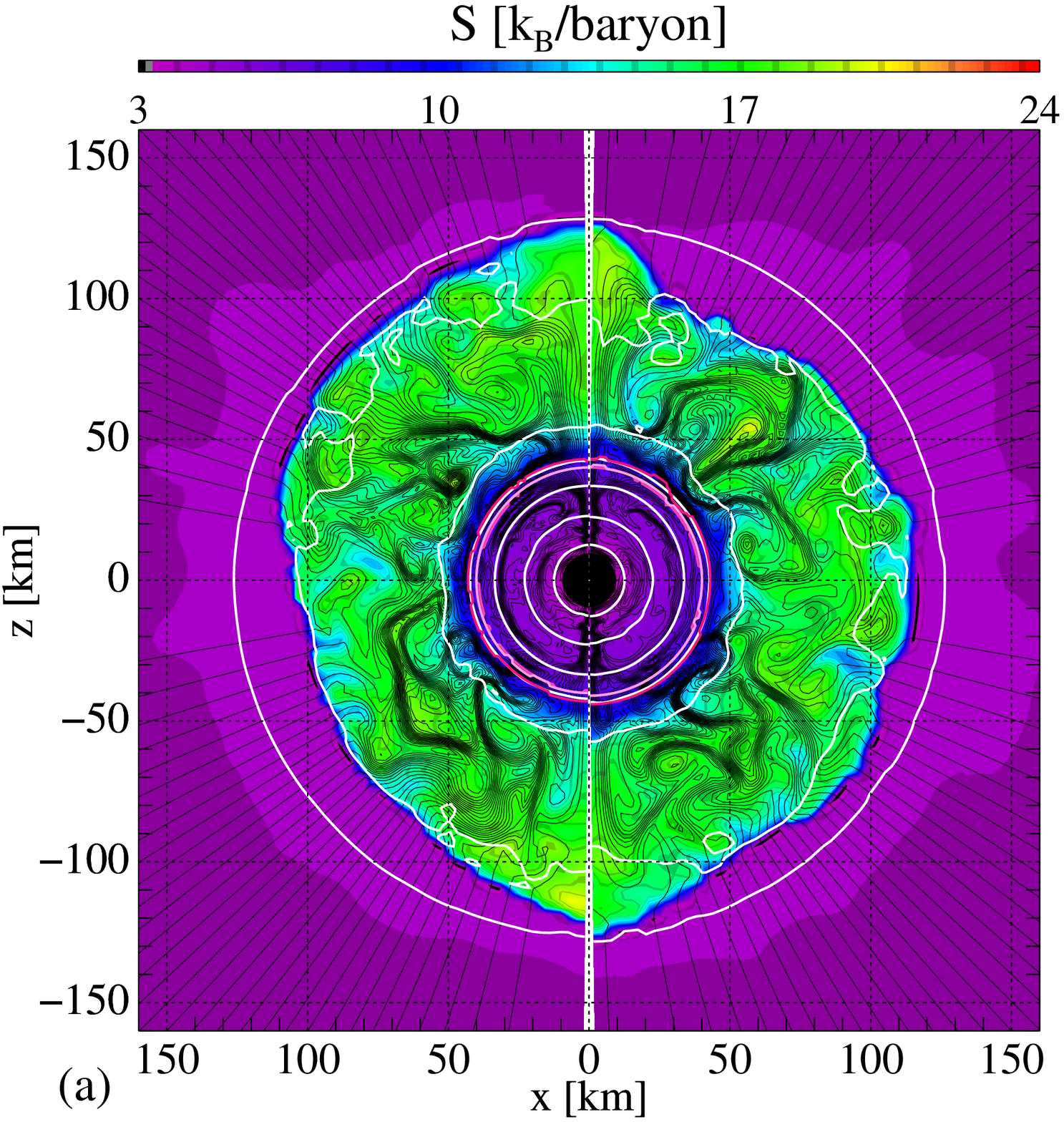}
  \includegraphics[width=0.49\linewidth]{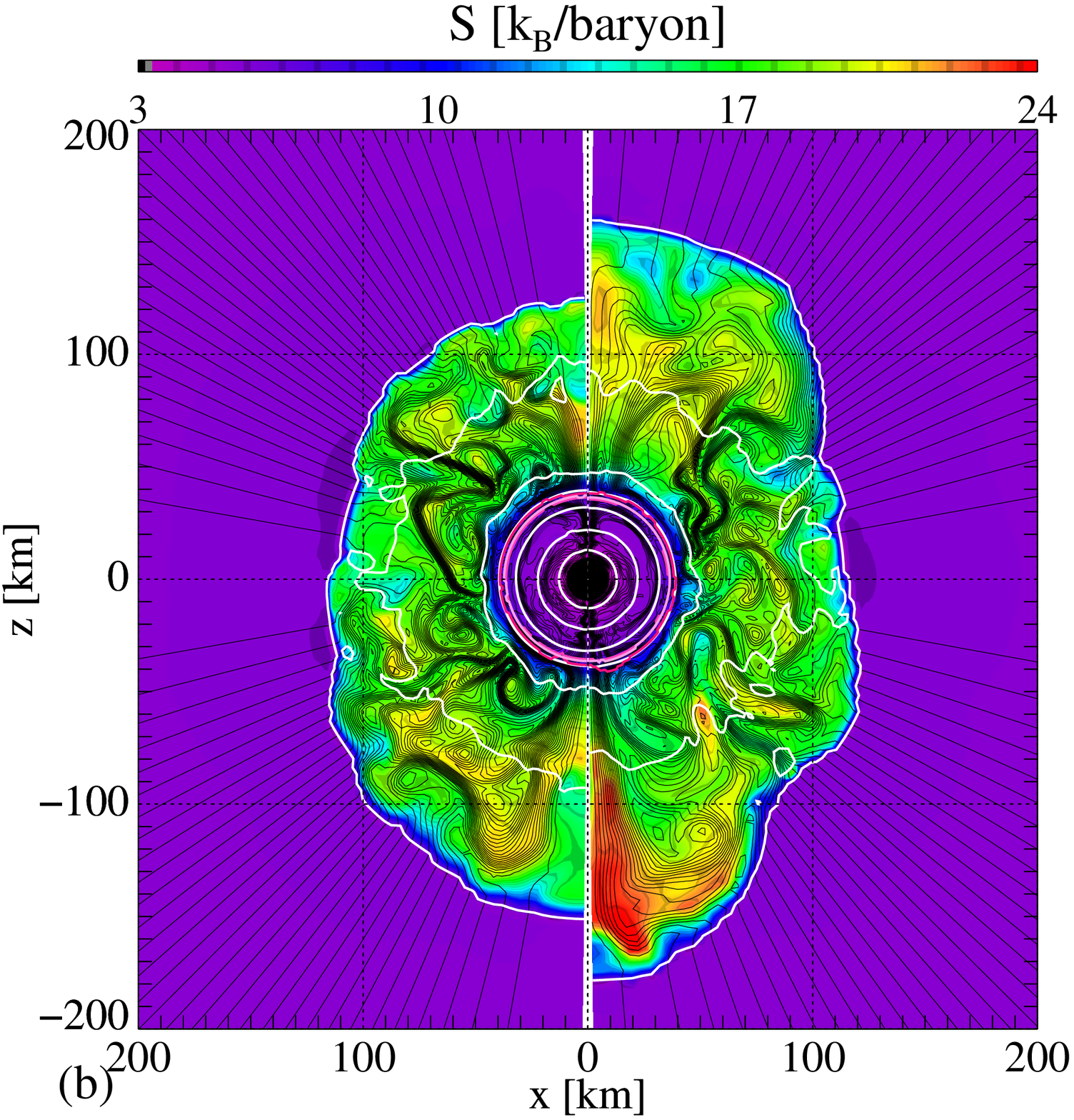}
  \caption{
    Maps of the specific entropy (colour scale) of
    \modls{s20-1} (left parts of the three panels) and
    \modelname{s20-2} (right parts).  The additional lines are
    magnetic field lines (thin black), density contours (white), and
    the three \nusps (pink).  The times of the snapshots are, from
    left to right, $t = 256$ (\panel{a}) and $296 \, \msec$ (\panel{b}).
  }
  \label{Fig:s20-2--2}
\end{figure*}

Around the time of shock revival, the conditions in \modl{s20-2}
undergo an important change: the magnetic pressure, before mostly
smaller than the gas pressure, achieves and exceeds thermal
equipartition.  The growth of the parameter $\beta^{-1} =
P_{\mathrm{mag}} / P_{\mathrm{gas}}$, shown in \panel{b} of \figref{Fig:s20-2--2}, to
values $\beta^{-1} \gtrsim 1$ coincides with the development of
outflows in \modl{s20-2}, whereas it does not happen in the less
magnetized \modl{s20-1}.  Hence, the structure of the polar region
cannot be understood without taking into account the magnetic field.

The field can be described in terms of flux tubes as displayed in
\figref{Fig:s20-2--2}.  Inside the \nusps (pink lines), the field is
concentrated between convective cells, in particular along the axis.
At the surface of the PNS, we find a layer of enhanced magnetic field
with, apart from the polar region, a strong $\theta$-component. This
layer is connected to the pre-shock field by various flux tubes
threading the gain layer, where they are advected, stretched, and
folded by the fluid flow.  The magnetic field is too weak to react
back on the flow for \modl{s20-1} at all times and for
\modelname{s20-2} before $t \approx 250 \, \msec$.  In these cases,
the dynamics of the flux tubes is highly stochastic with field lines
following the rise and fall of matter.  The result is a complex
pattern of field lines and the appearance, merger, and disruption of
hot bubbles such as the ones shown by the yellow and red colours in
\panel{b} \figref{Fig:s20-2--2} for \modl{s20-1}.

The super-equipartition field developing in \modl{s20-2} (see the
right part of \panel{b} of \figref{Fig:s20-2--1}) takes the form of a
rather thick radial flux tube around the axis.  In contrast to the
patterns at lower latitudes, this flux tube is maintained as a
coherent structure for a very long time.  The enhanced magnetic
pressure leads to a sideways expansion of the gas until the total, \ie
thermal plus magnetic, pressure is matched with the environment.
Consequently, the gas in the flux tube has a lower density than the
surroundings, as we see, \eg, in the white density contour bending
towards lower radii, $r \approx 70 \, \km$ at the axis in the right
part of \panel{b}.  Hence, buoyancy forces lead to the rise of gas
along the field lines.

Additionally, the reduction of gas density and internal energy causes
the gas to lose less energy by neutrino emission and shifts the
boundary between net heating and cooling towards lower radii.  Within
a few tens of ms after its onset, this process is effective along the
entire length of the radial flux tube, \ie starting immediately
outside the \nusp.  After $t \approx 300 \, \msec$, the
gain radius has receded to the surface of the PNS.  Hence, gas is
exposed to intense neutrino heating starting at very low radii as it
is ejected.  Furthermore, $\bar{\nu}_e$ are emitted with slightly
higher luminosities and mean energies than \mbox{$\nu_e$}. Therefore,
they are absorbed by the matter at higher rates.  Consequently, matter
is (re-)leptonized as it enters the region where it is injected into
the outflow.  This mode of launching the explosion generates an
outflow of high entropy.

\begin{figure}
  \centering
  \includegraphics[width=\linewidth]{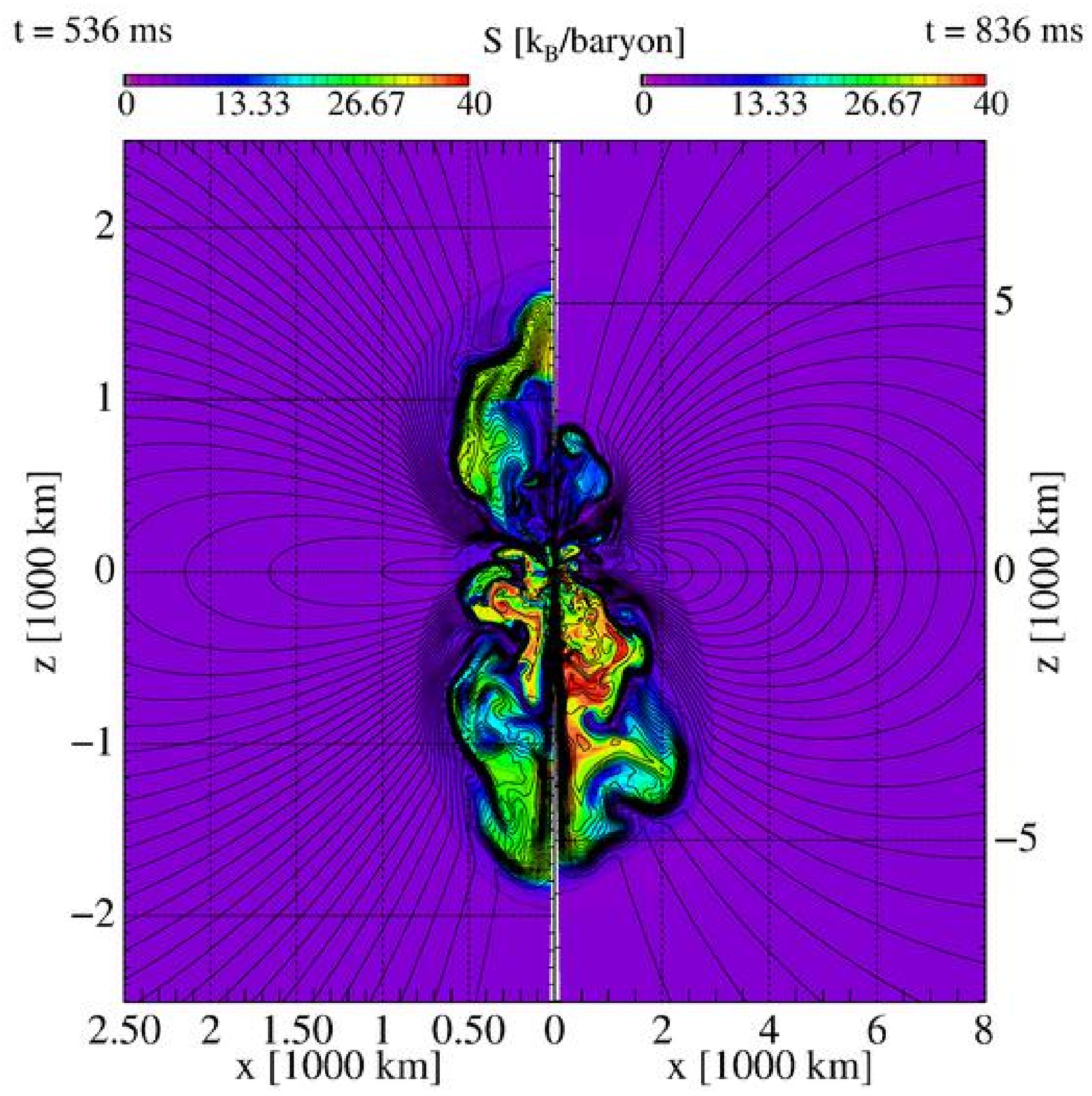}
  \caption{
    Maps of the specific entropy and magnetic field lines
    of \modl{s20-2} at two times long after the onset of the explosion
    as indicated in the panel.
  }
  \label{Fig:s20-2--3}
\end{figure}

The processes described here are at work at both poles and generate
outflows along the rotational axis between which accretion onto the
PNS proceeds (see the low-entropy regions close to
the equator in \figref{Fig:s20-2--3}).  The
accretion streams possess a stochastic nature and fall onto the PNS
with varying geometries and at different angles.  For most of the
time, they are situated close to the equator, but if one of them moves
to a pole, it may suppress the acceleration of the ejecta.  This
happens at the north pole at $t \approx 500 \, \msec$.  The system
then loses its approximate equatorial symmetry.  The southern outflow
continues progressing in the same way as before, while its northern
counterpart slows down as no more energy is injected at its base.
About 400 ms later, the north polar engine becomes active again.  At
$t\approx 835\,\msec$
(\figref{Fig:s20-2--3}, right part),
the outflows have reached radii of $r \approx 3000 \, \km$ and
$r \approx 5400 \, \km$ at the north and south poles, respectively.
By the end of the simulation, these values increase to
$r \approx 4000 \, \km$ and $r \approx 8400 \, \km$.

\subsection{\Modl{s20-3}}
\label{sSek:res:s20-3}

The evolution of \modl{s20-3} is affected critically by the rapid
rotation.  As we see in \panel{a} of \figref{Fig:s20-2--global},
the specific angular momentum of the core exceeds that found in
\modl{s20-2} by about an order of magnitude, rising to $j \gtrsim
10^{16} \, \cm^2 / \sec$ in the outer layers.  Compared to
\modl{s20-2}, $j$ has a more complex time evolution in the outer
layers with a pronounced dip around $t \approx 250 \, \msec$.  The
total rotational energy increases monotonically from $E^{\mathrm{rot}}
\approx \zehnh{2}{51} \, \erg$ shortly after bounce to
$E^{\mathrm{rot}} \approx \zehn{52} \, \erg$ at the end of the
simulation.

In the first few tens of milliseconds, the average magnetic field
rises very quickly in all layers of the PNS
(\figref{Fig:s20-2--global}, \panel{b}).  This rise is rather a
consequence of the rapid infall of magnetized matter and its
accumulation on the PNS than of amplification occurring in the PNS.
Immediately after bounce, the poloidal magnetic field
(\figref{Fig:s20-2--global}, \panel{c}) is far stronger than that of
\modl{s20-2}.  However, rather quickly the field growth ceases and, in
particular, in the innermost layers the field decreases temporarily
until $t \sim 300 \, \msec$.  The corresponding layers of \modl{s20-2}
amplify the field for a longer time with the consequence that the
initially weaker magnetized model develops a stronger field at layers
above $\rho \gtrsim 10^{13} \, \gccm$.  The premature stop of field
amplification and hence the inversion between the two models is caused
mainly by the magnetic field of \modl{s20-3} trying to enforce rigid
rotation in the central regions and slowing down the winding of
poloidal into toroidal field.  The field on the PNS surface, however,
decreases less strongly and exceeds $b \sim 10^{12} \, \Gauss$.  Panel
\banel{(c)} of \figref{Fig:s20-2--global} demonstrates that the
decrease of the total field strength and the partial inversion of the
ordering between \modls{s20-3} and \modelname{s20-2} only affect the
toroidal field.  The poloidal component, on the other hand, is
stronger for \modl{s20-3} than in the corresponding layers of
\modl{s20-2} at all times.

\Modl{s20-3} explodes considerably earlier than\modl{s20-2}.  The
shock begins to expand from a radius around $r_{\mathrm{sh}} \approx
170 \, \km$, which is similar to the stagnation radius of the other
models, at the south pole at $t \approx 100 \, \msec$ and about 15 ms
later also at the north pole (\cf \figref{Fig:global}).  The explosion
shock travels very fast.  Within a time of 300 ms, the shock wave
expands to a radius $r \approx 1000 \, \km$.  Its pattern speed is at
that point $v_{\mathrm{sh}} \approx 0.15 \, c$ and increases even
somewhat afterwards, and the flow speed reaches $v \approx 0.4 \, c$.
The mass and energy of the unbound ejecta are rising approximately
linearly after $t \approx 400 \, \msec$ at rates
$\dot{M}_{\mathrm{ej}} \sim 0.55 \msol/\sec$ and
$\dot{E}_{\mathrm{ej}} \sim \zehnh{2}{51} \, \erg/\sec$ and reach
values of $M_{\mathrm{ej}} \approx 0.45 \, \msol$ and $E_{\mathrm{ej}}
\approx \zehnh{2.1}{51} \, \erg / \sec$, respectively, at the end of
the simulation.  We note that the mass and energy are still rising at
that time and, hence, these values should be interpreted as lower
bounds on the final ones.  The same statement holds for the diagnostic
energy, which reaches $E_{\mathrm{exp}} \approx \zehnh{1.6}{51} \,
\erg$ at $t \approx 800 \, \msec$ and does not show any sign of
reducing the rate at which it grows.

After the explosion starts, the mass accretion rate across the shock,
thus far dropping at the same rate as in \modls{s20-1/2}, is reduced
even further.  Like in \modl{s20-2}, the explosion does not spell an
end to accretion as matter is still falling onto the PNS at low
latitudes.  Consequently, the PNS mass increases throughout the entire
simulation, although always remaining lower than in the other models.
The reduced accretion rate translates into lower neutrino emission.
After the accretion of the core interface at $t \approx 250 \, \msec$,
the luminosities of $\nu_e$ and $\bar{\nu}_e$ stabilize at slightly
more than half the corresponding values of the non-exploding
\modl{s20-1}, and those of the heavy flavours are almost as large as
the neutrinos of electron type.  We note, however, that the
luminosities are slightly below those of \modl{s20-1} even before the
onset of the explosion.  Although during this early phase, mass falls
through the shock wave at the same rate as in the other models, the
partial centrifugal support leads to a more aspherical shape of the
PNS with a higher equatorial radius.  As a consequence, the same mass
accretion rate corresponds to a lower release rate of gravitational
binding energy and, hence, less neutrino emission.  This effect
becomes more pronounced at later times when the minimum (\ie, polar)
radius of the PNS contracts similarly to the PNS radii of the other
two models, while the maximum (\ie, equatorial) radius shrinks much
slower and during a period of almost 150 ms starting at $t \approx 280
\, \msec$ even expands slowly (see \figref{Fig:global}, \panel{d}, for
$r_{\nu}$).  The axis ratio of the electron-\nusp assumes a peak of
$\approx 17:10 $ during this expansion phase to later settle to a
value around $13:10$.

As in the case of \modl{s20-2}, the global variables describing the
gain layer turn out to be insufficient for understanding the explosion
mechanism.  We find that in the run-up to the explosion, the mass and
the internal energy contained in the gain layer behave very similarly
to the other models and start to deviate only after the shock is
revived.  The neutrino heating rate is, in fact, smaller, leading to a
larger heating time scale (orange lines in \panel{c} and
\panel{d} of \figref{Fig:global-2}).  Equality between heating and
advection times is only reached once the shock has already travelled
to more than 1000 km.

The explosion, thus, is started at a time when the ram pressure of the
pre-shock matter is still too strong for shock revival in 
\modl{s20-2} even though the neutrino emission is reduced \wrt that
model.  Consequently, neutrino heating cannot be the most important
contribution to the explosion mechanism.

\begin{figure*}
  \centering
  \includegraphics[width=0.48\linewidth]{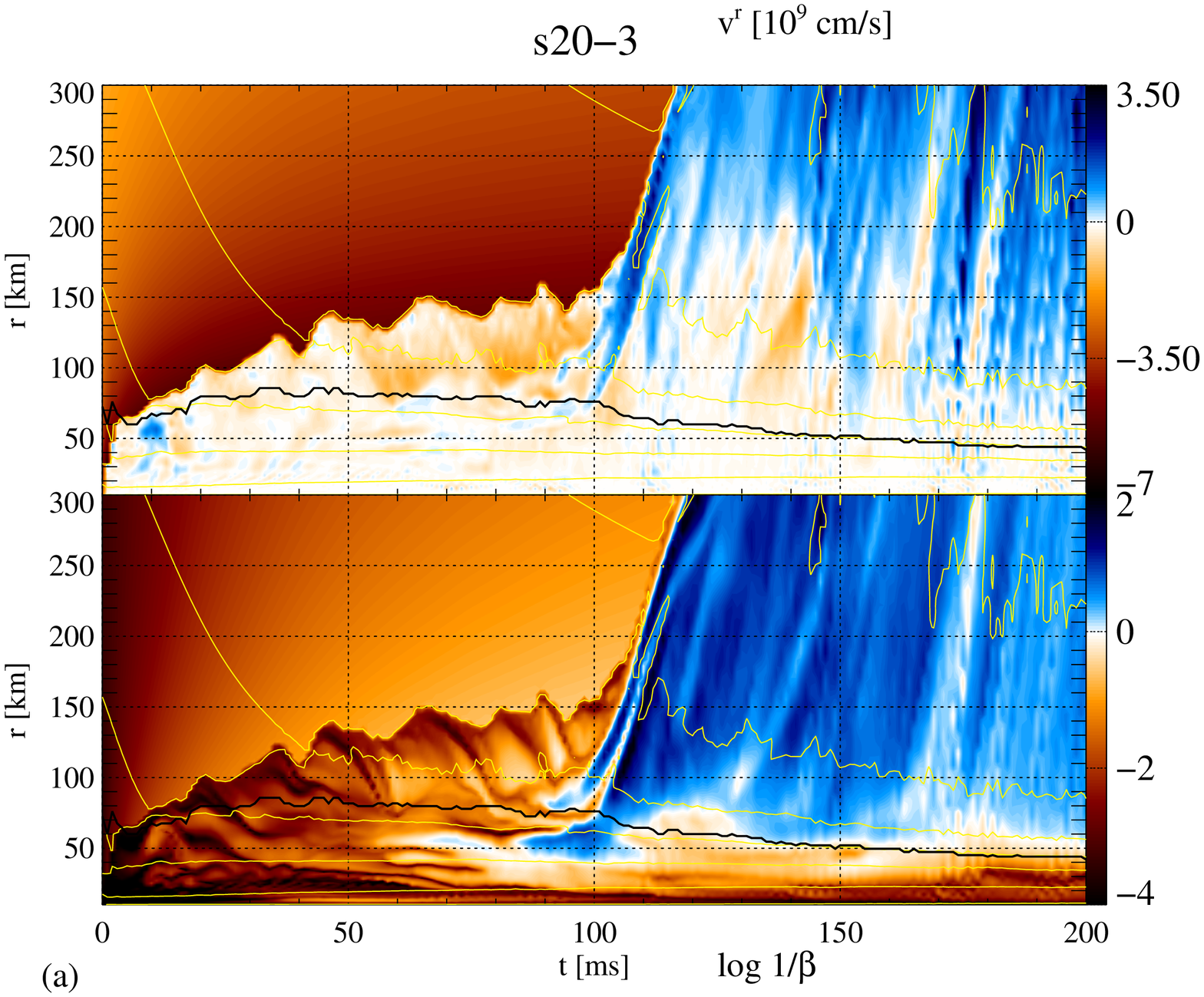}
  \includegraphics[width=0.48\linewidth]{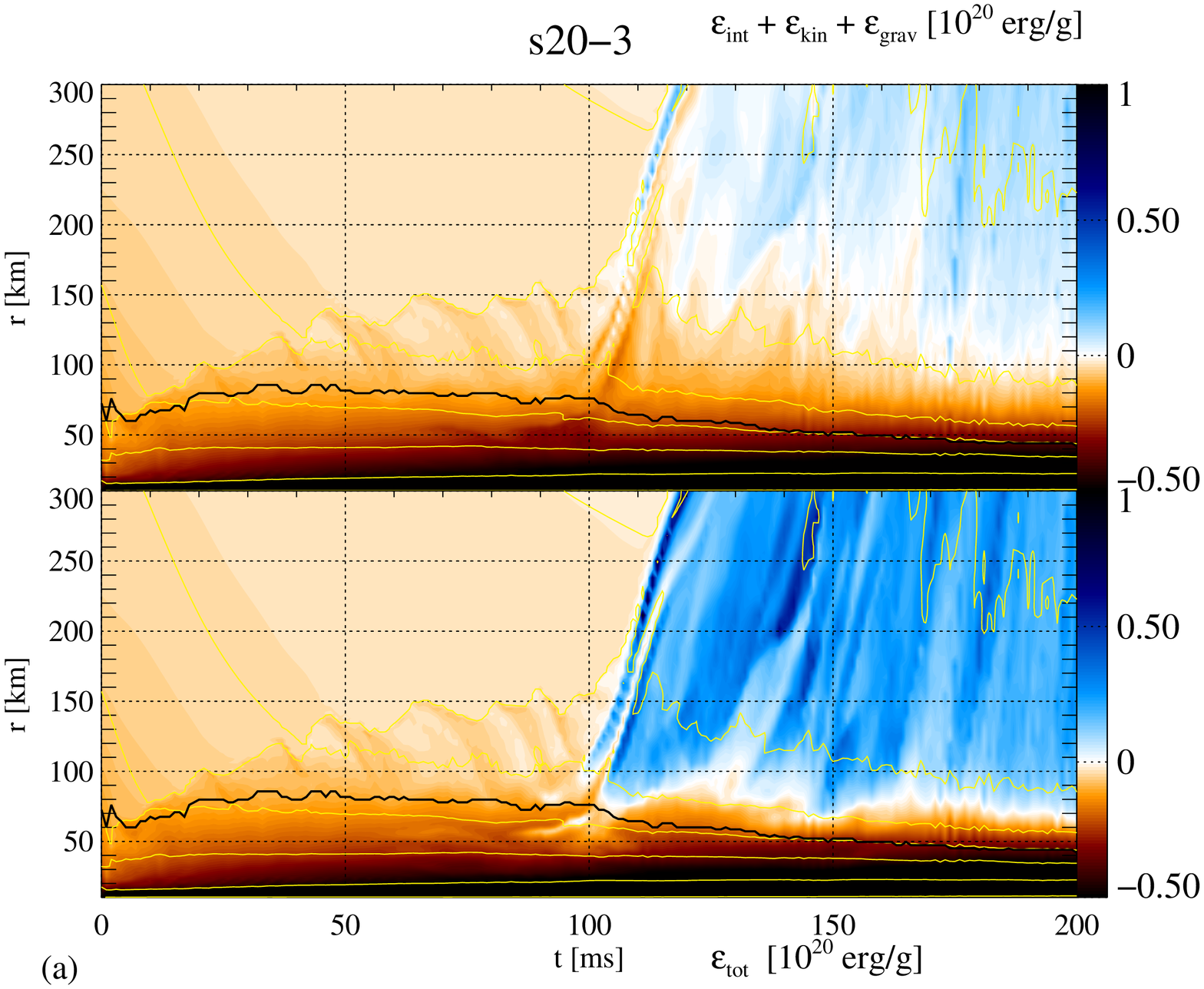}
  \caption{%%
    \Panel{a}: evolution of the radial velocity (top half) and
    the inverse plasma-$\beta$ parameter along the south pole of
    \modl{s20-3} (bottom half).
    \Panel{b}: the same for the specific total energy without (top) and
    with (bottom) the magnetic contribution.
    In both panels, the black lines represent the gain radius, while
    the yellow lines are iso-density contours.
  }
  \label{Fig:s20-3--1}
\end{figure*}

The shock runaway sets in with matter obtaining large positive
velocities (blue regions in the top part of \figref{Fig:s20-3--1}
\panel{a}).  The ejected matter is threaded by a mostly radial
magnetic field whose pressure exceeds the gas pressure by up to more
than one order of magnitude (bottom part of the same panel).  The
outward moving matter achieves positive total energies (blue regions
in \figref{Fig:s20-3--1} \panel{b}) close to the gain radius.  The
important contribution of the magnetic field to unbinding these fluid
elements is highlighted by a comparison between the total energy
including (bottom part of \panel{b} of \figref{Fig:s20-3--1}) or not
including (top part of the same panel), the magnetic energy
contribution.  The latter is in general significantly smaller as well
as positive only outside a much larger radius than the former.  This
difference strengthens the suggestion that the magnetic field, rather
than neutrino heating, is primarily responsible for the explosion.
The particular strength of the strong magnetic field in the polar
region is a result of the contraction of the magnetic flux creating a
very strong radial component oriented along the symmetry axis.  Other
effects such as differential rotation generating a toroidal component
or the MRI, which operates further inside the core, are not as
important.  The explosion mechanism bears some resemblence to the one
found by
\cite{Obergaulinger_et_al__2014__mnras__Magneticfieldamplificationandmagneticallysupportedexplosionsofcollapsingnon-rotatingstellarcores}
in non-rotating cores.  This similarity points towards a larger role
of the magnetic field than of the rotation in this shock revival.

\begin{figure*}
  \centering
  \includegraphics[width=0.49\linewidth]{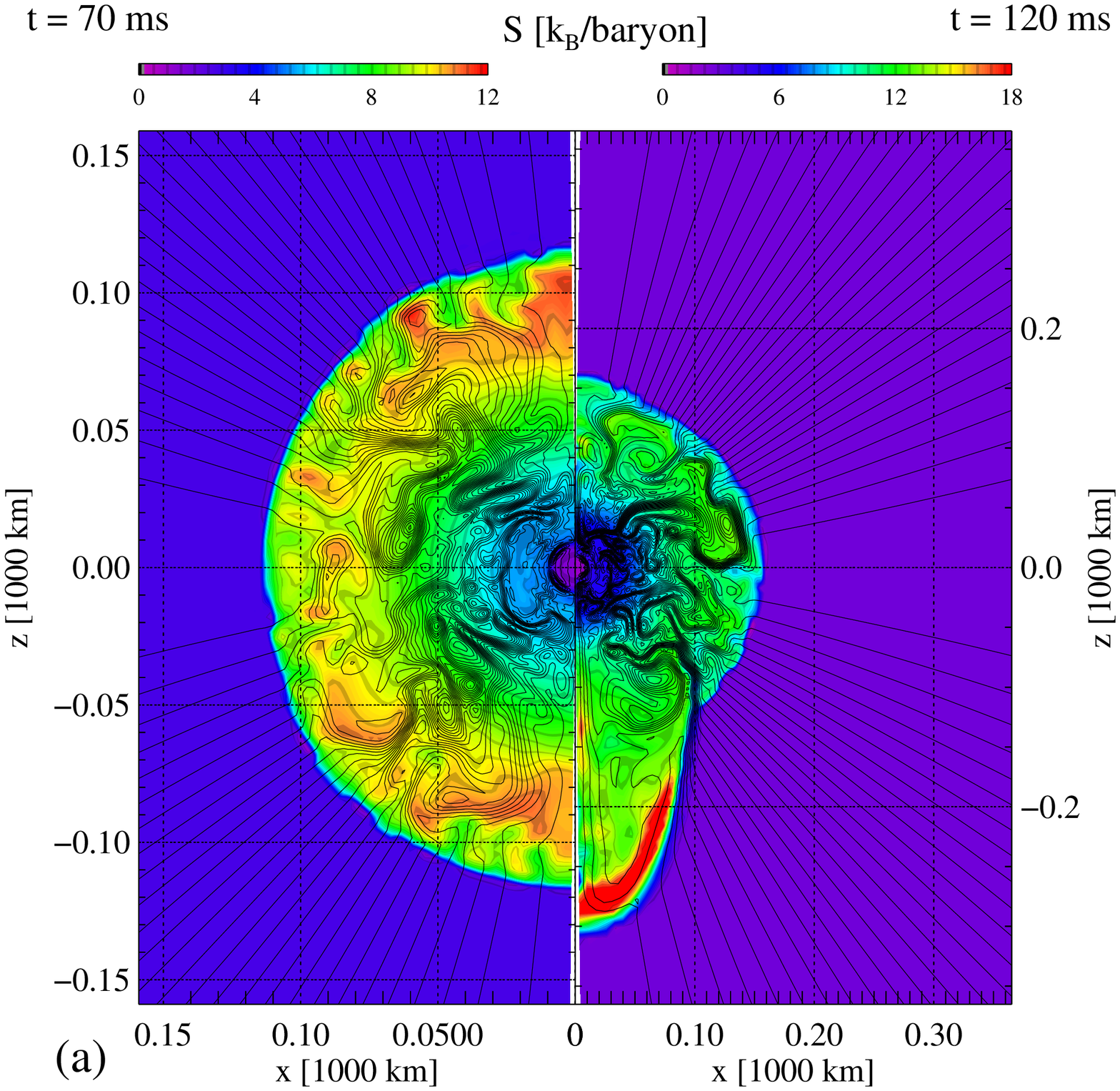}
  \includegraphics[width=0.49\linewidth]{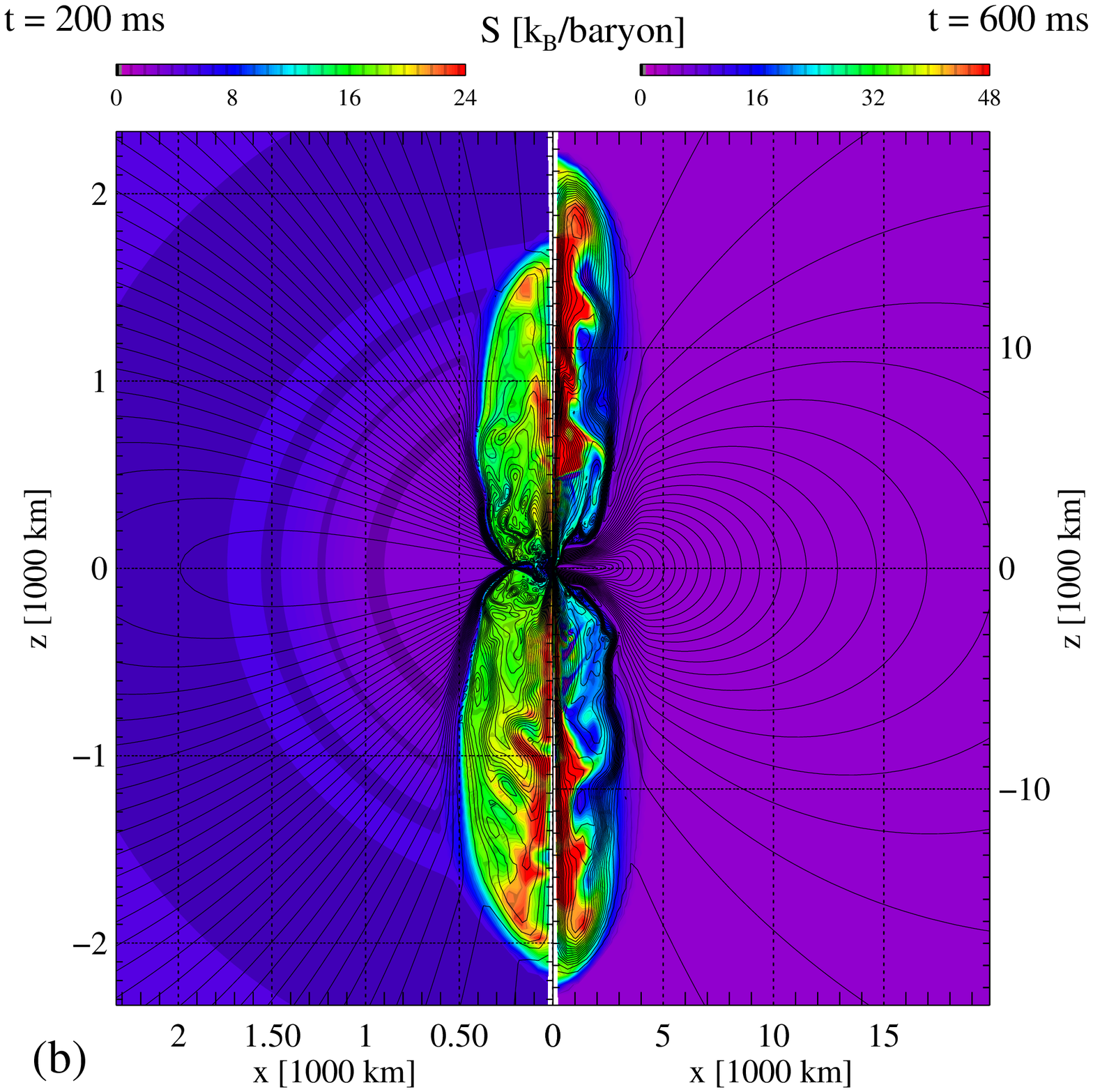}
  \caption{%%
    Maps of the specific entropy (colour
    scale) and field lines (black) of \modl{s20-3} for four different
    times as indicated in the panels.
  }
  \label{Fig:s20-3--2}
\end{figure*}

\begin{figure*}
  \centering
  \includegraphics[width=0.49\linewidth]{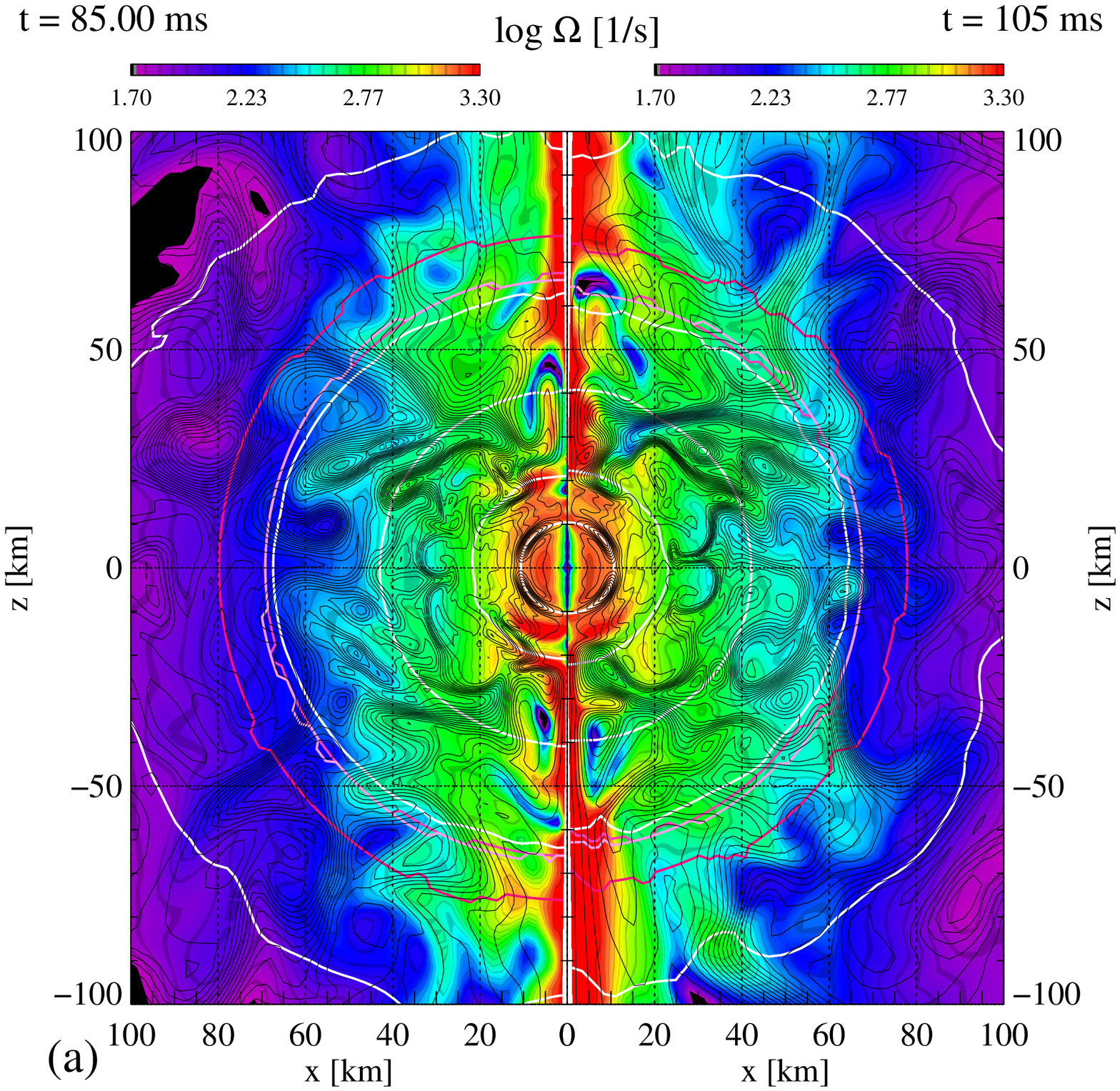}
  \includegraphics[width=0.49\linewidth]{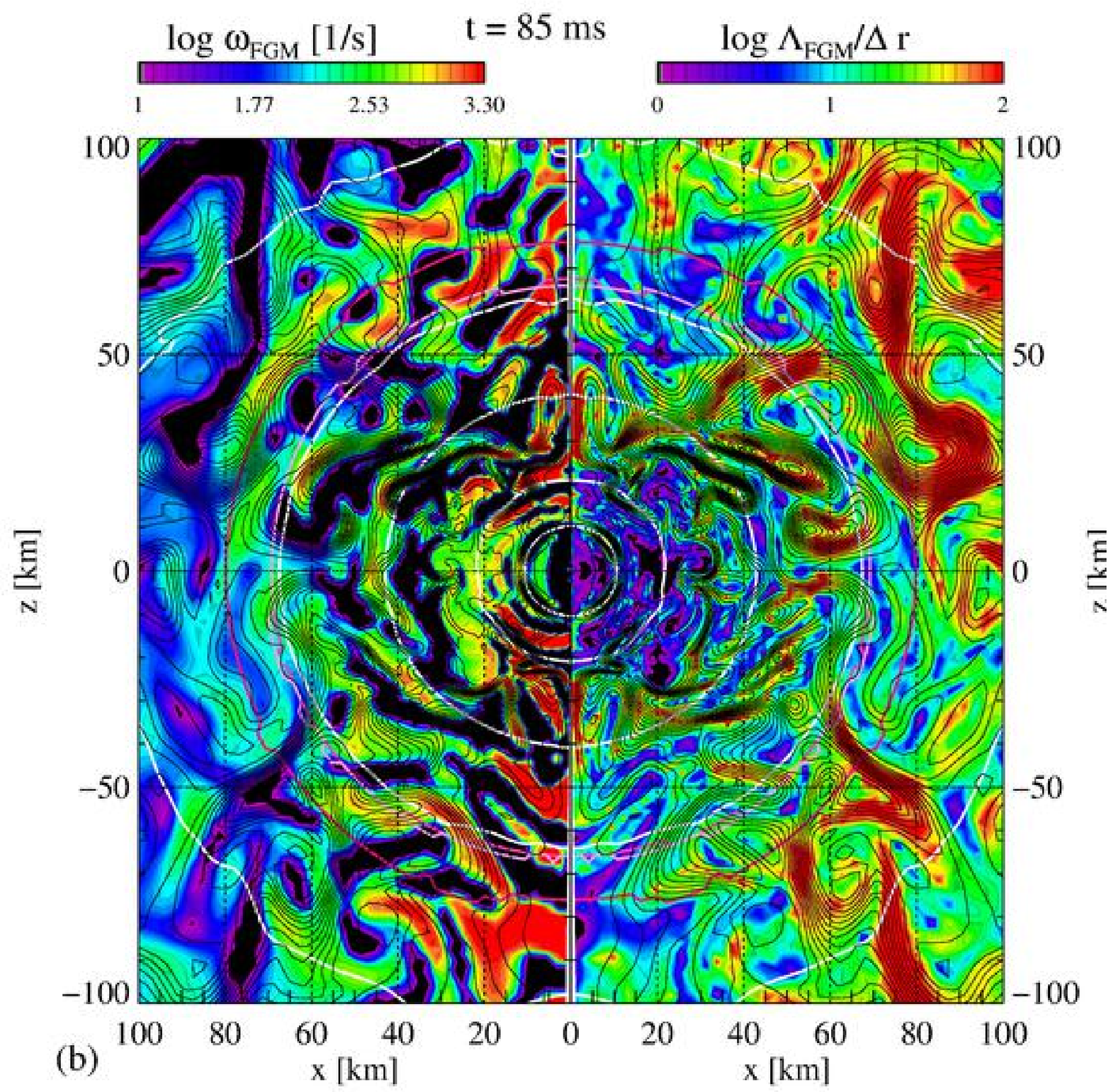}
  \caption{%%
    Panel \banel{(a)}: close-up on the PNS of \modl{s20-3} at the two
    indicated times.  The angular velocity is displayed in colours.
    Field lines are black, and white and pink lines are contours of
    density ($\rho = 10^{14,13,...} \, \gccm$) and the three \nusps,
    respectively.  Panel \banel{(b)}: growth rate,
      $\omega_{\mathrm{FGM}}$, and wave length,
      $\Lambda_{\mathrm{FGM}}$, of the fastest growing MRI mode
      relative to the grid width, $\Delta r$, in
      \modl{s20-3} at $t = 85 \, \msec$.
    }
  \label{Fig:s20-3--3}
\end{figure*}

The maps displayed in \panel{a} of \figref{Fig:s20-3--2} show the core
briefly before and after launching the explosion.  At $t \approx 70 \,
\msec$, the PNS is surrounded by a gain layer in which hydrodynamic
instabilities produce non-spherical flows and hot bubbles of
intermediate sizes appear for brief periods of time and where the
evolution and amplification of the magnetic field follows the dynamics
of the flow.  The feedback of the field on the flow, weak at most
locations throughout the gain layer, is important in several smaller
regions.  The first of those consists of strong sheets of
anti-parallel field lines form at intermediate latitudes at the PNS
surface.  They form, as we see in \panel{a} of \figref{Fig:s20-3--3},
at strong negative derivatives \wrt $\varpi$ of the angular velocity,
which inside the PNS has a roughly cylindrical profile.  The magnetic
field grows in these structures and leads to the transport of angular
momentum reflected in the distortion of the $\Omega$-profile.  Their
location, shape, and orientation suggest an interpretation as channel
modes of the magneto-rotational instability (MRI).  This
interpretation is consistent with the presence of strongly negative
radial gradients of the angular velocity, satisfying the criterion for
MRI growth. The latter, in its most basic form (for simplicity, we
only focus on the non-convective magneto-shear modes of
\cite{Obergaulinger_etal__2009__AA__Semi-global_MRI_CCSN}) predicts
that the MRI grows in regions fulfilling the condition
$\mathcal{R_{\varpi}} = \varpi \partial_{\varpi} \Omega^2 < 0$
($\varpi$ is the cylindrical radius).  The growth rate of the fastest
growing modes of the MRI, given by $\omega_{\mathrm{FGM}} = -
\mathcal{R}_{\mathrm{\varpi}} / (4\Omega)$, is shown in the left part
of panel \banel{(b)} of \figref{Fig:s20-3--3} at $t = 85 \, \msec$.
While the values vary strongly across the core, the predicted growth
rate in many locations is very high with $\omega_{\mathrm{FGM}}$ in
the range of at least several $100 \, \sec^{-1}$.  On average, lower
radii correspond to higher growth rates.  Such high values agree with
the exponential growth of the (poloidal) magnetic field in the density
range $10^{11} \, \gccm < \rho < 10^{13} \, \gccm$ (panels
\banel{(b,c)} of \figref{Fig:s20-2--global}).  As a demonstration, we
added an exponential function with a growth rate of $\omega = 100 \,
\sec^{-1}$ (solid black line) to the curve showing the growth of the
poloidal field strength in the density range $10^{12} \, \gccm < \rho
< 10^{13} \, \gccm$.  Furthermore, our grid is sufficiently fine to
resolve the fastest growing MRI modes, whose length scales are several
km.  The right part of panel \banel{(b)} of \figref{Fig:s20-3--3}
depicts the variable $\Lambda_{\mathrm{FGM}} = 2 \pi
c_{\mathrm{A}}^{\mathrm{pol}}/\sqrt{\mathrm{abs}(-\mathcal{R}_{\mathrm{\varpi}})}$
which is equal to theoretical wavelength of the fastest growing mode,
$\lambda_{\mathrm{FGM}} = 2 \pi
c_{\mathrm{A}}^{\mathrm{pol}}/\sqrt{-\mathcal{R}_{\mathrm{\varpi}}}$
in unstable regions, in units of the local grid width, $\Delta r$,
where $c_{\mathrm{A}}^{\mathrm{pol}}$ is the Alfv\'en speed
corresponding to the poloidal magnetic field component.  In the
regions of fastest MRI growth, the wavelength typically exceeds
several km, which is in general comparable to or larger than the local
grid width.  We note, however, that there are regions where the
$\lambda_{\mathrm{MRI}}$ is under-resolved.  The MRI is better
resolved at larger radii.  There, however, the growth rate is rather
low.  We typically cover $\lambda_{\mathrm{FGM}}$ by a few grid cells,
which, together with geometrical effects such as the orientation of
the field \wrt the gradient of $\Omega$ and the neglect of the thermal
stratification, accounts for a reduction of the numerical growth rate
\wrt the theoretical predictions.  We note that the distributions of
both the growth rate and the wavelength bear the imprint of the
channel modes that at this time have already developed and modified
the rotational profile.  Despite the simplifications entering our
analysis, the 2d maps in \figref{Fig:s20-3--3} support the likely
possibility that the MRI is resolved and operating in \modl{s20-3}, at
least initially in the form of channel modes.

After their formation, the interior end points of these channel modes
tend to converge at the polar axis as, \eg, at $t = 105 \, \msec$
around $z = \pm 40 \, \km$ (\figref{Fig:s20-3--2}, right part of
\panel{a}).  This convergence further enhances the quite strong field
that is already there and generates a strong increase of the magnetic
pressure.  The magnetic pressure is strong enough to overcome the ram
pressure of the surrounding matter and finally drives a polar outflow
containing strongly magnetized gas (\cf\,the distribution of the
specific entropy at $t = 120 \, \msec$ in the \panel{a} of
\figref{Fig:s20-3--2}).

The magnetic forces accelerate matter above the PNS surface into
collimated outflows during the entire run of the simulation.  The rate
at which they transfer energy to the gas is sustained at values above
$10^{50} \erg / \sek$ and is highly variable, leading to a rich
substructure of faster and slower fluid elements.  We find at $t = 200
\, \msec$ (\figref{Fig:s20-3--2}, \panel{b}) regions of high entropy
along the axis of the outflow, which 400 ms later have intensified.
At that later time, we also note the formation of reverse shocks
situated at $z \sim + 4000 \, \km$ separating hot matter outside from
cooler gas to the interior (visible as the rear end of the red region
in the right part of the panel).

The region above the poles of the PNS continues to inject mass from an
equatorial accretion stream that separates the north and south
outflows into the polar outflows.  However, by the end of the
simulation, its mass, $M_{\mathrm{PNS}} \approx 1.93 \, \msol$, is
still far from the threshold for instability against collapse to a
black hole.  Based on the mass accretion rates at that point and
taking into account the centrifugal stabilization, we estimate that
the PNS can maintain its stability for several seconds, if not
indefinitely.

\section{Summary and conclusions}
\label{Sek:SumCon}

Increasingly sophisticated multidimensional core collapse simulations
of large sets of stars with state-of-the-art neutrino transport are
getting closer to an explanation of the mechanisms that produce
successful CCSN explosions or non-exploding cores resulting in
collapse to black holes. However, a small fraction of all progenitors
may possess strong magnetic fields that cannot be ignored as is
commonly done in the later simulations, in particular when combined
with rapid rotation.  However, studies of magneto-rotational core
collapse have to face the problem that one-dimensional stellar
evolution modelling only provides limited information on the magnetic
field and rotation of the pre-collapse cores.  Therefore, profiles of
magnetic field and rotational velocity are commonly assumed for cores
that are otherwise evolved to the pre-collapse state without magnetic
fields and rotation.  Thus, these studies were able to identify
various MHD processes that may in general play a role on the explosion
dynamics, but could only in a few cases arrive at detailed conclusions
for specific progenitors.

Lacking stellar evolution results for magnetized, rotating versions of
the star under consideration, we worked in the same framework to study
the influence of rotation and magnetic fields on the evolution of the
core of a star of an initial mass of $M = 20 \, \msol$ by means of
special relativistic axisymmetric MHD simulations including an
approximately relativistic gravitational potential and a spectral
two-moment neutrino transport.  We compared three models with the same
distribution of density, electron fraction, and temperature, but
different strength of the magnetic field and different angular
velocities.  

The first of our models, \modelname{s20-1}, starts with negligible
rotation and magnetic field.  This model behaves essentially as if
there were no rotation and magnetic fields at all.  About 100 ms after
core bounce, the shock wave stagnates before beginning to contract
gradually.  No explosion is achieved, even though conditions
for shock
revival, as measured in the ratio between the time scale for advection
of fluid elements through the gain layer and the time scale for
neutrino heating, improve when the ram pressure at the shock decreases
after the accretion of the surface of the iron core.  Though the
initial magnetic field is amplified both in the PNS and in the gain
layer by convection and the SASI, it never reaches a strength
sufficient to affect the evolution.

We computed \modl{s20-2}, a version of the model with a $j$-constant
rotational law, a central angular velocity of $\Omega_0 = 0.1 \,
\sek^{-1}$ and the same magnetic fields as \modl{s20-1}, \viz
$B_0^{\mathrm{p,\phi}} = 10^{10,11} \, \Gauss$ for the poloidal and
toroidal components, respectively.  Compared to estimates from stellar
evolution theory, the field strength is strong, but not exceedingly
so.  The specific angular momentum increases with radius to values of
$j \sim 10^{15} \, \cm^2 \, \sek^{-1}$, \ie lies below the values of
$j \sim 3 \times 10^{16} \, \cm^2 \, \sek^{-1}$ considered the
threshold for the formation of a GRB central engine within the collapsar model
\cite{MacFadyen_Woosley__1999__ApJ__Collapsar}.

The influence of rotation and magnetic fields lead to a deviation of
the evolution of \modl{s20-2} from \modl{s20-1} after the accretion of
the silicon shell.  The contraction of the shock wave is stopped and a
bipolar explosion is launched.  Rapid, albeit subrelativistic,
outflows develop along the rotational axis while downflows continue at
lower latitudes.  Neutrino heating, though certainly playing its role
in the explosion mechanism, does not differ significantly from the
non-exploding \modl{s20-1}, as both the total luminosities and heating
rates are very similar in both models.  The fact that \modl{s20-2}
develops an explosion at a neutrino luminosity, which is insufficient
in \modelname{s20-1}, is consistent with the findings of
\cite{Iwakami_et_al__2014__apj__ParametricStudyofFlowPatternsbehindtheStandingAccretionShockWaveforCore-CollapseSupernovae}
who show that rotation tends to reduce the critical luminosity for
shock revival.  Striking differences at the local level can explain
the different evolution of the models.  The outflows are launched from
a region of strongly enhanced magnetic field, whose pressure becomes
equal to and even greater than the gas pressure, close to the \nusp of
the core.  Magnetic buoyancy in a thick radial flux tube accelerates
gas radially along the polar axis.  The explosion found in this model
is, at least during the simulated time, of moderately high energy: the
diagnostic explosion energy reaches about $E_{\mathrm{exp}} \approx
\zehnh{4.6}{50} \, \erg$, but can be expected to increase further.

\Modl{s20-3}, with ten times faster rotation and a ten times stronger
poloidal magnetic field than, but the same toroidal field as,
\modl{s20-2}, explodes already at $t \approx 100 \, \msec$.  The
explosion shows little contribution by neutrino heating.  In fact, the
rapid rotation reduces the neutrino luminosity, which makes a
neutrino-driven explosion more unlikely (\cf
\cite{Summa_et_al__2017__ArXive-prints__Rotation-supportedNeutrino-drivenSupernovaExplosionsinThreeDimensionsandtheCriticalLuminosityCondition}).
Nevertheless, the model is able to overcome the high ram pressure of
the infalling matter during the accretion of the iron core due to a
strong, super-equipartition field along the rotational axis.  Beyond
compression, hydrodynamic instabilities, and winding of poloidal
field, we also find field amplification by the MRI in this model in
the form of a very rapid rise of the field strength produced by the
geometrical convergence of MRI channel modes along the rotational axis
and close to the surface of the PNS.  The properties of these modes as
found in the simulation are in reasonable agreement with the
theoretical predictions for the magneto-shear modes of the MRI.  Our
numerical grid covers each wavelength (in the region of interest,
typically a several km) by a few grid cells.  This means the MRI is
resolved, though in the regions of fastest growth only fairly
marginally.  In the weaker magnetized and slower rotating
\modl{s20-2}, the grid would also suffice to resolve the MRI.  There,
however, the growth rates are much lower, and any possible MRI
activity would be hidden behind the complex dynamics dominated by
other effects.

In \modl{s20-3}, the rotation leads to a notable flattening of the PNS
with a final equator-to-pole axis ratio of $13:10$, while the spin of
\modl{s20-2} is too slow to affect the shape significantly.  Both PNSs
rotate differentially at the end of the simulation.  Over a large
fraction of the PNS volume, the angular velocity profile is
cylindrical.  The magnetic field decreases from the center to the
surface by about 2 to 3 orders of magnitude in strength.  In both
cases, it is dominated by the toroidal component.  The poloidal
components possess average surface strengths between a few times
$10^{11} \, \Gauss$ (\modelname{s20-2}) and $10^{12} \, \Gauss$
(\modelname{s20-3}).  Both components show a rich substructure and
cannot adequately be described by a simple low-order multipole, in
line with the findings of
\cite{Obergaulinger_Aloy__2017__JournalofPhysicsConferenceSeries__Evolutionofthesurfacemagneticfieldofrotatingproto-neutronstars}.

We find that \modl{s20-3} explodes very energetically.  Its diagnostic
energy grows to $E_{\mathrm{exp}} \approx \zehnh{1.6}{51} \, \erg$ at
the end of the simulation.  The final value is most likely
considerably higher as we find still a roughly linear increase of
$E_{\mathrm{exp}}$ when the simulation was ended.  Such a high energy
is similar to values computed by
\cite{Burrows_etal__2007__ApJ__MHD-SN} for magnetorotational
explosions.

We close by briefly discussing the limitations of our present study.
As noted above, the lack of detailed stellar-evolution models with
rotation and magnetic fields forces us to artificially add those two
components to a spherically symmetric pre-collapse model.  Thus, our
study belongs to those investigating the fundamental effects of
rotation and magnetic fields rather than those able to make
predictions for a specific star.  As such, the models are only a small
part of the possible conditions and a more comprehensive scan of the
parameter space might be in order.  As far as the simulations are
concerned, the most crucial limitation is certainly the assumption of
axisymmetry.  This restriction makes a difference in the tendency of
models to explode without rotation and magnetic fields, but also with
rotation only
\cite{Summa_et_al__2017__ArXive-prints__Rotation-supportedNeutrino-drivenSupernovaExplosionsinThreeDimensionsandtheCriticalLuminosityCondition}.
If magnetic fields are included, the impact is even larger.  Hence, a
continuation of the study in a three-dimensional setup is required.

\section*{Acknowledgements}
\label{Sek:Ackno}

MO and MAA acknowledge support from the European Research Council
(grant CAMAP-259276) and from the Spanish Ministry of Economy and
Finance and the Valencian Community grants under grants
AYA2015-66899-C2-1-P and PROMETEOII/2014-069, resp.  OJ acknowledges
support by the Special Postdoc Researcher (SPDR) program and iTHEMS
cluster of RIKEN as well as the European Research Council through
grant ERC-AdG No. 341157-COCO2CASA.  We thank Thomas Janka for
valuable help, in particular regarding the aspects of microphysics and
neutrinos.  The computations were performed under grants
AECT-2016-1-0008, AECT-2016-2-0012, AECT-2016-3-0005,
AECT-2017-1-0013, AECT-2-0006, AECT-2017-3-0007, and AECT-2018-1-0010
of the Spanish Supercomputing Network on clusters \textit{Pirineus} of
the Consorci de Serveis Universitaris de Catalunya (CSUC),
\textit{Picasso} of the Universidad de M{\'a}laga,
\textit{FinisTerrae} of the Centro de Supercomputaci{\'o}n de Galicia
(CESGA), and \textit{MareNostrum} of the Barcelona Supercomputing
Centre, respectively, and on the clusters \textit{Tirant} and
\textit{Lluisvives} of the Servei d'Inform\`atica of the University of
Valencia. We thank the PHAROS COST Action (CA16214) and the GWverse
COST Action (CA16104) for partial support.

\end{document}